\documentstyle[amssymb,pra,aps,psfig]{revtex}

\newcommand{\be}{\begin{equation}}
\newcommand{\ee}{\end{equation}}
\newcommand{\br}{\begin{eqnarray}}
\newcommand{\er}{\end{eqnarray}}

\newcommand{\bd}{\begin{displaymath}}
\newcommand{\ed}{\end{displaymath}}

\newcommand{\bfig}{\begin{figure}}
\newcommand{\efig}{\end{figure}}

\def\3cdot{\cdot \cdot \cdot}

\def\om0{\omega _0}
\def\Om0{\Omega _0}

\def\text#1{{\rm{#1}}}

\def\->{\rightarrow}
\def\=>{\Rightarrow}
\def\-->{\longrightarrow}
\def\==>{\Longrightarrow}

\def\pr{^\prime}
\def\pr2{^{\prime\prime}}

\def\bfig{\begin{figure}}
\def\efig{\end{figure}}

\begin{document}
\title{Decoherence in strongly coupled quantum oscillators}
\author{M. A. de Ponte$^{1}$, M. C. de Oliveira$^{2}$, and M. H. Y. Moussa$%
^{3}$}
\address{Departamento de F\'{\i}sica, CCET, Universidade Federal de S\~{a}o\\
Carlos, Via Washington Luiz Km 235, S\~{a}o Carlos, 13565-905, SP, Brazil.}
\maketitle

\begin{abstract}
In this paper we present a comprehensive analysis of the coherence
phenomenon of two coupled dissipative oscillators. The action of a classical
driving field on one of the oscillators is also analyzed. Master equations
are derived for both regimes of weakly and strongly interacting oscillators
from which interesting results arise concerning the coherence properties of
the joint and the reduced system states. The strong coupling regime is
required to achieve a large frequency shift of the oscillator normal modes,
making it possible to explore the whole profile of the spectral density of
the reservoirs. We show how the decoherence process may be controlled by
shifting the normal mode frequencies to regions of small spectral density of
the reservoirs. Different spectral densities of the reservoirs are
considered and their effects on the decoherence process are analyzed. For
oscillators with different damping rates, we show that the worse-quality
system is improved and vice-versa, a result which could be useful for
quantum state protection. State recurrence and swap dynamics are analyzed as
well as their roles in delaying the decoherence process.
\end{abstract}


\section{Introduction}

The process of decoherence of quantum states has long been a central issue
in the description of quantum measurements \cite{Neumann,Zurek,CL}. In
recent years, experimental advances in the domain of cavity QED and trapped
ions have allowed the decoherence of photon \cite{Haroche} and phonon \cite%
{Wineland} field states to be probed in more depth, providing insights into
the borderline between the classical and quantum descriptions of the
physical world. The decoherence time of a superposition of coherent states
in a cavity field was measured \cite{Haroche} and shown to be in full
agreement with theoretical predictions \cite{Glauber,WM}. In trapped ions
systems, the observed damping of Rabi oscillations has motivated a number of
articles on the main sources of noise leading to decoherence \cite%
{Milburn,Vogel,Serra,Budini}. Such experimental achievements in matter-field
interactions have also encouraged a deep dialog between theoretical and
experimental physics, resulting in a degree of mastery of fundamental
quantum phenomena that may herald a new stage in the technology of
communication \cite{Comm} and computation \cite{Comp}.

The exploration of the borderline between quantum and classical descriptions
of nature \cite{Zurek} has impelled the generation of superposition states
of mesoscopic systems, known as \textquotedblleft Schr\"{o}dinger cat
states\textquotedblright\ \cite{Haroche,Wineland}. Such superpositions are
irreversibly affected by their surroundings, whose effect is to destroy
probability interference (coherence), and continuously transformed into
statistical mixtures. Thus, the environment plays a key role in the
establishment of a direct correspondence between quantum and classical
dynamics. While the decoherence time of a superposition state depends on the
amplitude of the field, the relaxation does not, since the model adopted for
the relaxation process is amplitude damping, achieved by coupling the
systems bilinearly to the degrees of freedom of the reservoir.

Decoherence and its dependence upon the amplitude of the superposition state
is the main obstacle to the implementation of a logic network based on
quantum gates \cite{nielsen,cnot}. The dream of quantum communication and
computation comes up against the nightmare of decoherence mechanisms \cite%
{Haroche1}, owing not only to the inevitable action of the surrounding
environment but also to the intrinsic fluctuations in the interaction
parameters required for logic operations \cite{Milburn,Vogel,Serra}. The
need for huge superpositions of qubit states in the practical implementation
of logical operations imposes the requirements that the quantum systems be
totally isolated from the environment and that the interaction parameters
involved be tightly controlled. For this reason, investigation of the
sources of noise in such promising quantum systems is a crucial step towards
the realization of a quantum logic processor. There is also a major effort
being made in present-day research, to discover mechanisms to prevent
decoherence occurring in actual physical systems, by using parity kicks \cite%
{Vitali}, stroboscopic feedback \cite{Tombesi}, engineered driving fields %
\cite{pump1} or an engineered reservoir \cite{Zoller,Wineland1,Matos}. In
this light, the main concern of the present work is to analyze the coherence
dynamics and decoherence process in a network composed of two coupled
dissipative oscillators, which may be field modes in dissipative cavities %
\cite{Raimond}, phonon modes of trapped ions \cite{Ions}, phonon modes of
surface electrons in liquid helium \cite{Nelson}, etc. Master equations are
derived for both weakly and strongly interacting oscillators, leading to
interesting results concerning the coherence properties of the joint and the
reduced system states. This work constitutes a first step towards a more
comprehensive treatment of the decoherence process in multipartite quantum
systems.

On attempting to extend the work on decoherence to interacting quantum
oscillators coupled to distinct reservoirs, one faces the problem of
deriving a master equation for different regimes of coupling between
oscillators. Leaving to one side the difficulty of engineering an arbitrary
coupling strength between the oscillators, in the present work we analyze
not only the weak, but also the strong coupling regime, where the coupling
strength between the oscillators is near the typical oscillators
frequencies. In both regimes we assume that the coupling strength between
the oscillators is considerably larger than the system damping rates.

In weak coupling, the coupling strength between the oscillators, labelled $%
\ell =1,2$ from here on, is considerably smaller than the typical
frequencies of either oscillator, and the resulting master equation is as if
the two oscillators were decoupled and a decay channel, described by the
Liouville operator ${\cal L}_{\ell }\rho $, can simply be inserted into the
master equation for each oscillator considered. In that case, assuming both
oscillators have the same damping constant, the decoherence time for each
oscillator is unaffected by the interaction with the other one. However,
when the oscillators have different damping constants (field modes in
cavities with different quality factors, for example), we observe that the
``good-quality''\ oscillator gets worse, while the ``bad-quality''\
oscillator gets better, a result which can be employed for quantum state
protection.

In the strong coupling regime, we observe that a cross-decay channel ${\cal L%
}_{12}\rho $ appears, besides the usual system-reservoir individual decay
channels ${\cal L}_{\ell }\rho $. This cross-decay channel modifies the
decoherence process of both the joint and the reduced system state, to an
extent depending crucially on the spectral density of the reservoirs. In
fact, in the strong coupling regime, the normal-mode frequencies are
substantially shifted from the typical oscillator frequencies, enabling us
to explore the whole profile of the spectral densities of the reservoirs. We
show how the decoherence process may be controlled by shifting the
normal-mode frequencies to regions of small spectral density of the
reservoirs. Apart from these spectral densities, the competition between the
cross-decay and the usual channels can give rise to a computed delay or
advance of the decoherence process, for eigenstates of the system normal
modes. Note that if a system pointer variable does not commute with the
operator responsible for its coupling to other system, it is clear that the
internal dynamics must interfere in its decoherence time. Thus the
derivation of master equations for strongly interacting systems is a central
task \cite{Walls,Charmicael} in the study of decoherence in quantum networks.

It is worth mentioning some previous work concerned with coupled systems. In
Ref. \cite{Raimond} the authors describe a proposal to achieve reversible
decoherence of a mesoscopic superposition of field states. This proposal is
based on the possibility of performing a reversible coupling between two
Fabry-Perot cavities. In Ref. \cite{Nemes}, a theoretical model of the
experimental proposal in Ref. \cite{Raimond} is given, but in \cite{Nemes}
the inevitable coupling of the resonators to their environment is taken into
account when the reversibility of coherence loss is analyzed. A system of
two coupled cavities is also analyzed in Ref. \cite{Zoubi}, where just one
of the cavities is interacting with a reservoir. In Ref. \cite{Zoubi}, a
master equation is derived in the case of strongly coupled cavities and it
is shown that the relaxation term is not simply the standard one, obtained
by neglecting the interaction between the cavities. It is the aim of the
present paper, in the context of cavity QED, to analyze the reversible
decoherence process of Refs. \cite{Raimond,Nemes}, where two dissipative
cavities are considered, but investigating also the regime of strongly
coupled cavities, as done in \cite{Zoubi}, in which some remarkable
coherence properties appear. A central result extracted from our discussion
is that in a strongly interacting quantum network the decoherence time may
not decrease as the number of systems considered grows. In fact, it may
happen that with many coupled sites \cite{Net}, as with the two coupled
systems analyzed here, the decoherence time increases, depending on the
spectral density of the reservoirs.

Together with the strong coupling between two oscillators we consider a
classical driving field feeding one of the oscillators continuously, which
is intended to drive each of the coupled oscillators to a stationary
coherent state \cite{pump1,MMC}, other than the vacuum state. We select a
specific coupling between the oscillators, which may be responsible for the
dynamics of local transfer of states or state swap between them, as
discussed in Ref. \cite{meu} in relation to cavity QED with weak coupling.

The outline of this paper is as follows. In Sec. II we develop a master
equation for the two coupled lossy oscillators, one of which is under the
action of a classical driving field, and we analyze the weak and strong
field coupling regimes. As the spectral density of the reservoirs plays a
crucial role in the strong coupling regime, in Sec. III we analyze
particular cases of spectral densities. In Sec. IV we develop a c-number
version of the master equation and solve it with reservoir temperatures set
to zero. The solution of the master equation derived for strongly coupled
oscillators is also analyzed in Sec. IV for special initial field states. In
Sec. V we discuss the recurrence and swapping dynamics of the system states.
The central result of the paper, namely the coherence properties of the
system states, which depend on the spectral density of the reservoir and are
strongly affected by the regime of coupling, is presented in Sec. VI. In
Sec. VII we consider the two oscillators to have different damping rates and
demonstrate that the coupling between them, assumed to be larger than these
damping rates, makes the good-quality oscillator worse and the bad-quality
oscillator better. A careful analysis of the entropy excess in our network
is developed in Sec. VIII. Finally, Sec. IX concludes the paper.

\section{The problem of coupled dissipative oscillators: derivation of the
master equation}

General results can be extracted from specific examples of quantum
oscillators, such as field modes in coupled cavities, trapped ions, or
surface electrons in liquid helium. Let us consider a system of two
interacting oscillators under the action of a driving field, as pictured in
Fig. 1. We start from a positive-defined Hamiltonian so that the energy
spectrum has a lower bound which is equal to zero in the absence of the
driving field \cite{FLO,Amir}. The system Hamiltonian is then given by
\begin{eqnarray}
H &=&\hbar \omega _{10}\left( a_{1}^{\dagger }+\frac{\lambda }{2\omega _{10}}%
a_{2}^{\dagger }\right) \left( a_{1}+\frac{\lambda }{2\omega _{10}}%
a_{2}\right) +\hbar \omega _{20}\left( a_{2}^{\dagger }+\frac{\lambda }{%
2\omega _{20}}a_{1}^{\dagger }\right) \left( a_{2}+\frac{\lambda }{2\omega
_{20}}a_{1}\right)  \nonumber \\
&&+\hbar \sum_{k}\omega _{1k}\left( b_{1k}^{\dagger }+\frac{V_{1k}}{\omega
_{1k}}a_{1}^{\dagger }\right) \left( b_{1k}+\frac{V_{1k}}{\omega _{1k}}%
a_{1}\right) +\hbar \sum_{k}\omega _{2k}\left( b_{2k}^{\dagger }+\frac{V_{2k}%
}{\omega _{2k}}a_{2}^{\dagger }\right) \left( b_{2k}+\frac{V_{2k}}{\omega
_{2k}}a_{2}\right)  \nonumber \\
&&+\hbar F\left( e^{i\omega t}+a_{2}^{\dagger }\right) \left( e^{-i\omega
t}+a_{2}\right) ,  \label{Eq1}
\end{eqnarray}%
where $a_{\ell }^{\dagger }$ and $a_{\ell }$ are, respectively, the creation
and annihilation operators for the oscillator mode of frequency $\omega
_{\ell 0}$, whereas $b_{\ell k}$ and $b_{\ell k}^{\dagger }$ are the
analogous operators for the $k$th bath mode of oscillator $\ell $, whose
corresponding frequency and coupling strength are $\omega _{\ell k}$ and $%
V_{\ell k}$, respectively. The coupling strength between the oscillators is $%
\lambda $ and the classical driving field applied to oscillator
$2$ has intensity $F$ and frequency $\omega $. Assuming that the
coupling between the oscillators and their reservoirs satisfies
the condition $\sum_{k}\left( V_{\ell k}\right) ^{2}/\omega _{\ell
k}\ll \omega _{\ell 0}$, \ and shifting the origin of the energy
scale to $\hbar F$, we obtain from Eq. (\ref{Eq1}) the Hamiltonian
$H=\sum_{\ell }H_{\ell }+H_{I}$, given by
\begin{mathletters}
\begin{eqnarray}
H_{\ell } &=&\hbar \omega _{\ell }a_{\ell }^{\dagger }a_{\ell }+\hbar
\sum_{k}\omega _{\ell k}b_{\ell k}^{\dagger }b_{\ell k}+\hbar
\sum_{k}V_{\ell k}\left( a_{\ell }b_{\ell k}^{\dagger }+a_{\ell }^{\dagger
}b_{\ell k}\right) {\rm {,}}  \label{Eq2a} \\
&&+\hbar F\left( a_{2}^{\dagger }%
\mathop{\rm e}%
\nolimits^{-i\omega t}+a_{2}%
\mathop{\rm e}%
{}^{i\omega t}\right) \delta _{\ell 2}{\rm {,}}  \nonumber \\
H_{I} &=&\hbar \lambda \left( a_{1}a_{2}^{\dagger }+a_{1}^{\dagger
}a_{2}\right) {\rm {.}}  \label{Eq2b}
\end{eqnarray}%
Here, $\omega _{\ell }$ is related to the natural frequency $\omega _{\ell
0} $ by
\end{mathletters}
\begin{equation}
\omega _{\ell }=\omega _{\ell 0}\left( 1+\frac{\lambda ^{2}}{4\omega
_{10}\omega _{20}}+\delta _{\ell 2}\frac{F}{\omega _{20}}\right) {\rm {,}}
\label{EqF}
\end{equation}%
such that in the limit of weak coupling between the oscillators ($\lambda
/\omega _{\ell 0}\ll 1$) and weak amplification process ($F/\omega _{20}\ll
1 $) we obtain the natural frequencies $\omega _{\ell }=\omega _{\ell 0}$.
In this limit, we could have started from Hamiltonian $H$ given by Eqs. (\ref%
{Eq2a}) and (\ref{Eq2b}) instead of (\ref{Eq1}). Note that since
we are assuming weak couplings between the oscillators and their
reservoirs, it is unnecessary to write these interactions in a
positive-defined form, as done in Eq. (\ref{Eq1}). However, the
positive-defined form for the coupling between the oscillators
ensures an energy spectrum with a lower bound (equal to zero when
the driving field is switched off), whatever the value of the
coupling $\lambda $. Under a unitary transformation represented by
the operator
\begin{equation}
U(t)=\exp \left[ -i\omega t\sum_{\ell }\left( a_{\ell }^{\dagger }a_{\ell
}+\sum_{k}b_{\ell k}^{\dagger }b_{\ell k}\right) \right] ,  \label{Eq3}
\end{equation}%
we obtain the transformed time-independent Hamiltonian
\begin{equation}
{\cal H}=U^{\dagger }(t)HU(t)+i\frac{dU^{\dagger }(t)}{dt}U(t)=\sum_{\ell }%
{\cal H}_{\ell }+{\cal H}_{I}{\rm {,}}  \label{Eq4}
\end{equation}%
where
\begin{mathletters}
\begin{eqnarray}
{\cal H}_{\ell } &=&\hbar \omega _{\ell }^{\prime }a_{\ell }^{\dagger
}a_{\ell }+\hbar \sum_{k}\omega _{\ell k}^{\prime }b_{\ell k}^{\dagger
}b_{\ell k}+\hbar \lambda \left( a_{1}a_{2}^{\dagger }+a_{1}^{\dagger
}a_{2}\right) {\rm {,}}  \label{Eq5a} \\
{\cal H}_{I} &=&\hbar \sum_{k}V_{\ell k}\left( a_{\ell }b_{\ell k}^{\dagger
}+a_{\ell }^{\dagger }b_{\ell k}\right) +\hbar F\left( a_{2}^{\dagger
}+a_{2}\right) \delta _{\ell 2}{\rm {,}}  \label{Eq5b}
\end{eqnarray}%
and the shifted frequencies are given by
\end{mathletters}
\begin{mathletters}
\begin{eqnarray}
\omega _{\ell }^{\prime } &=&\omega _{\ell }-\omega {\rm {,}}  \label{Eq6a}
\\
\omega _{\ell k}^{\prime } &=&\omega _{\ell k}-\omega {\rm {.}}  \label{Eq6b}
\end{eqnarray}

\qquad\ From here on we consider the specific case where $\omega
_{1}^{\prime }=\omega _{2}^{\prime }=\Omega $ (or $\omega _{1}=\omega _{2}$%
), which links the amplitude of the driving field with the natural
frequencies, as $F=(\omega _{10}-\omega _{20})\left[ 1+\lambda ^{2}/\left(
4\omega _{10}\omega _{20}\right) \right] $. Therefore, in the absence of the
driving field, the condition $\omega _{1}^{\prime }=\omega _{2}^{\prime
}=\Omega $, implies that $\omega _{10}=\omega _{20}$. With the condition $%
\omega _{1}^{\prime }=\omega _{2}^{\prime }=\Omega $, the Hamiltonian in Eq.
(\ref{Eq5a}) can be diagonalized through the canonical transformation
\end{mathletters}
\begin{mathletters}
\begin{eqnarray}
A_{1} &=&\frac{1}{\sqrt{2}}\left( a_{1}+a_{2}\right) ,  \label{Eq7a} \\
A_{2} &=&\frac{1}{\sqrt{2}}\left( a_{1}-a_{2}\right) {\rm {,}}  \label{Eq7b}
\end{eqnarray}%
where $A_{1}$ and $A_{2}$ satisfy the same commutation relation as $a_{1}$
and $a_{2}$: $[A_{i},A_{j}]=0$ and$\left[ A_{i},A_{j}^{\dagger }\right]
=\delta _{ij}$. The purpose of these new operators is to decouple the direct
interaction between oscillators $1$ and $2$, described by $\hbar \lambda
\left( a_{1}a_{2}^{\dagger }+a_{1}^{\dagger }a_{2}\right) $. Consequently,
indirect interactions between oscillators $1$ and $2$ will be created
through their respective reservoirs, as described by Hamiltonian ${\bf H}=%
{\bf H}_{0}+{\bf H}_{I}$, where

\end{mathletters}
\begin{mathletters}
\begin{eqnarray}
{\bf H}_{0} &=&\hbar \sum_{\ell }\left[ \Omega _{\ell }A_{\ell }^{\dagger
}A_{\ell }-(-1)^{\ell }\frac{F}{\sqrt{2}}\left( A_{\ell }^{\dagger }+A_{\ell
}\right) +\sum_{k}\omega _{\ell k}^{\prime }b_{\ell k}^{\dagger }b_{\ell k}%
\right] ,  \label{Eq8a} \\
{\bf H}_{I} &=&\frac{\hbar }{\sqrt{2}}\sum_{\ell ,k}\left[ V_{1k}\left(
A_{\ell }b_{1k}^{\dagger }+A_{\ell }^{\dagger }b_{1k}\right) -(-1)^{\ell
}V_{2k}\left( A_{\ell }b_{2k}^{\dagger }+A_{\ell }^{\dagger }b_{2k}\right) %
\right] ,  \label{Eq8b}
\end{eqnarray}%
and the frequencies are given by $\Omega _{\ell }=\Omega -(-1)^{\ell
}\lambda $. Without direct coupling between oscillators $1$ and $2$, as
modelled by Hamiltonian ${\bf H}$, it becomes simpler to derive the master
equation, following the reasoning in Ref. \cite{Walls}. In the interaction
picture, to the second order of perturbation, the evolution of the density
matrix of the coupled oscillators is given by
\end{mathletters}
\begin{equation}
\frac{d\rho _{12}(t)}{dt}=-\frac{1}{\hbar ^{2}}\int_{0}^{t}dt^{\prime }{\rm {%
Tr}}_{R}\left[ V(t),\left[ V(t^{\prime }),\rho _{R}(0)\otimes \rho _{12}(t)%
\right] \right] ,  \label{Eq9}
\end{equation}%
where $V(t)=\exp \left( i{\bf H}_{0}t/\hbar \right) {\bf H}_{I}\exp \left( -i%
{\bf H}_{0}t/\hbar \right) $. Note that the density matrix in the
interaction picture, $\rho _{12}(t)$, follows from the state vector
transformed by both unitary operators: $U^{\dagger }(t)$ defined by Eq. (\ref%
{Eq3}) and $\exp \left( i{\bf H}_{0}t/\hbar \right) $. Defining the operator
${\cal O}_{\ell }^{\dagger }(t)=\sum_{k}V_{\ell k}b_{\ell k}^{\dagger }\exp
\left( i\omega _{\ell k}^{\prime }t\right) /\sqrt{2}$, we proceed to obtain
the master equation, assuming that the reservoir frequencies are very
closely spaced, to allow a continuum summation. We have to solve the
integrals appearing in Eq. (\ref{Eq9}), related to correlation functions of
the form
\begin{eqnarray}
\int_{0}^{t}dt^{\prime }\left\langle {\cal O}_{\ell }^{\dagger }(t){\cal O}%
_{\ell }(t^{\prime })\right\rangle
\mathop{\rm e}%
\nolimits^{-i\Omega _{m}t+i\Omega _{n}t^{\prime }} &=&\int_{0}^{t}dt^{\prime
}\int_{0}^{\infty }\frac{d\omega _{\ell k}}{4\pi }V_{\ell }^{2}\left( \omega
_{\ell k}\right) \sigma _{\ell }^{2}\left( \omega _{\ell k}\right) (N_{\ell
}\left( \omega _{\ell k}\right) +1)  \nonumber \\
&&\times
\mathop{\rm e}%
\nolimits^{i(\omega _{\ell k}-\Omega _{n}-\omega )(t-t^{\prime })}%
\mathop{\rm e}%
\nolimits^{-i(\Omega _{m}-\Omega _{n})t}{\rm {,}}  \label{Eq10}
\end{eqnarray}%
where, from here on, $m,n=1,2$, the function $N_{\ell }\left( \omega _{\ell
k}\right) $ is defined by

\begin{equation}
\left\langle b_{\ell }^{\dagger }(\omega _{\ell k})b_{\ell }(\omega _{\ell
k^{\prime }})\right\rangle =2\pi N_{\ell }\left( \omega _{\ell k}\right)
\delta \left( \omega _{\ell k}-\omega _{\ell k^{\prime }}\right)
\label{Eq11}
\end{equation}%
and $\sigma _{\ell }\left( \omega _{\ell k}\right) $ is the density of
states of reservoir $\ell $. Performing the variable transformations $\tau
=t-t^{\prime }$ and $\varepsilon =\omega _{\ell k}-\Omega _{n}-\omega $, we
obtain, for $n=1$ and $2$, respectively,
\begin{mathletters}
\begin{eqnarray}
\int_{0}^{t}dt^{\prime }\left\langle {\cal O}_{\ell }^{\dagger }(t){\cal O}%
_{\ell }(t^{\prime })\right\rangle
\mathop{\rm e}%
\nolimits^{-i\Omega _{m}t+i\Omega _{1}t^{\prime }} &=&\frac{1}{4}%
\mathop{\rm e}%
{}^{-i(\Omega _{m}-\Omega _{1})t}\int_{-\omega _{\ell }^{+}}^{\infty
}d\varepsilon V_{\ell }^{2}\left( \varepsilon +\omega _{\ell }^{+}\right)
\sigma _{\ell }^{2}\left( \varepsilon +\omega _{\ell }^{+}\right)  \nonumber
\\
&&\times \left[ N_{\ell }\left( \varepsilon +\omega _{\ell }^{+}\right) +1%
\right] \int_{0}^{t}d\tau
\mathop{\rm e}%
\nolimits^{i\varepsilon \tau },  \label{Eq12a} \\
\int_{0}^{t}dt^{\prime }\left\langle {\cal O}_{\ell }^{\dagger }(t){\cal O}%
_{\ell }(t^{\prime })\right\rangle
\mathop{\rm e}%
\nolimits^{-i\Omega _{m}t+i\Omega _{2}t^{\prime }} &=&\frac{1}{4}%
\mathop{\rm e}%
{}^{-i(\Omega _{m}-\Omega _{2})t}\int_{-\omega _{\ell }^{-}}^{\infty
}d\varepsilon V_{\ell }^{2}\left( \varepsilon +\omega _{\ell }^{-}\right)
\sigma _{\ell }^{2}\left( \varepsilon +\omega _{\ell }^{-}\right)  \nonumber
\\
&&\times \left[ N_{\ell }\left( \varepsilon +\omega _{\ell }^{-}\right) +1%
\right] \int_{0}^{t}d\tau
\mathop{\rm e}%
\nolimits^{i\varepsilon \tau },  \label{Eq12b}
\end{eqnarray}%
where the frequency $\omega _{\ell }$ has been split into two effective
frequencies corresponding to the normal modes of the coupled oscillators
(note that $\omega _{1}=\omega _{2}$):
\end{mathletters}
\begin{equation}
\omega _{\ell }^{\pm }=\omega _{\ell }\pm \lambda .  \label{Eq13}
\end{equation}%
We have suppressed the contribution of the Cauchy principal value since it
represents only a small shift in the frequency $\omega _{\ell }^{\pm }$. We
note that the minimum value of $\omega _{\ell }^{-}$ is $F\left[ \omega
_{10}/\left( \omega _{10}+\omega _{20}\right) \right] $ which follows from $%
\lambda =2\omega _{20}$. Next, we discuss both regimes: $(i)$ the weak
coupling regime $\lambda /\omega _{\ell 0}\ll 1$, and $(ii)$ the strong
coupling regime $\lambda /\omega _{\ell 0}\approx 1$. For the strong
coupling regime we will assume $\lambda /\omega _{\ell 0}=2$ in order to
minimize $\omega _{\ell }^{-}$, which becomes zero when the driving field is
switched off.

As usual, we consider that $V_{\ell }$, $\sigma _{\ell }$,\ and $N_{\ell }$\
are functions that vary slowly around the frequency $\omega _{\ell }^{\pm }$%
, an assumption which does not apply to the function $N_{\ell }\left( \omega
_{\ell }^{-}\right) =\left[ \exp \left( \hbar \omega _{\ell }^{-}/kT\right)
-1\right] ^{-1}$ (taking the reservoir to be in thermal equilibrium at
temperature $T$) in the strong coupling regime when $F=0$, since in this
case $\omega _{\ell }^{-}\approx 0$. However, this regime (even with $F=0$)
can safely be applied to a reservoir at absolute zero, the situation we
analyze in the present work. We observe that, in practice, $N_{\ell }\left(
\omega _{\ell }^{-}\right) \approx 0$ whenever the shift in the frequency $%
\omega _{\ell }^{-}$, arising from the contribution of the Cauchy principal
value, becomes sufficiently greater than $kT/\hbar $. Note that the last
integrals in Eqs. (\ref{Eq12a}) and (\ref{Eq12b}) contribute significantly
only when $|\varepsilon t|\lessapprox 1$, so that if we extend the upper
limit of the time integration to infinity, the expressions for the
correlation functions become
\begin{mathletters}
\begin{eqnarray}
\int_{0}^{t}dt^{\prime }\left\langle {\cal O}_{\ell }^{\dagger }(t){\cal O}%
_{\ell }(t^{\prime })\right\rangle
\mathop{\rm e}%
\nolimits^{-i\Omega _{m}t+i\Omega _{1}t^{\prime }} &=&\frac{\gamma _{\ell
}(\omega _{\ell }^{+})}{2}\left[ N_{\ell }\left( \omega _{\ell }^{+}\right)
+1\right]
\mathop{\rm e}%
{}^{-i(\Omega _{m}-\Omega _{1})t},  \label{Eq14a} \\
\int_{0}^{t}dt^{\prime }\left\langle {\cal O}_{\ell }^{\dagger }(t){\cal O}%
_{\ell }(t^{\prime })\right\rangle
\mathop{\rm e}%
\nolimits^{-i\Omega _{m}t+i\Omega _{2}t^{\prime }} &=&\frac{\gamma _{\ell
}(\omega _{\ell }^{-})}{2}\left[ N_{\ell }\left( \omega _{\ell }^{-}\right)
+1\right]
\mathop{\rm e}%
{}^{-i(\Omega _{m}-\Omega _{2})t},  \label{Eq14b}
\end{eqnarray}%
where the damping rates are defined as
\end{mathletters}
\begin{equation}
\gamma _{\ell }(\omega _{\ell }^{\pm })=\frac{1}{2}V_{\ell }^{2}\left(
\omega _{\ell }^{\pm }\right) \sigma _{\ell }^{2}\left( \omega _{\ell }^{\pm
}\right) \int_{-\omega _{\ell }^{\pm }}^{\infty }d\varepsilon \delta
(\varepsilon ).  \label{Eq15}
\end{equation}

Defining $\Gamma _{\ell }=$ $V_{\ell }^{2}\left( \omega _{\ell 0}\right)
\sigma _{\ell }^{2}\left( \omega _{\ell 0}\right) $, in the weak coupling
regime, where $\omega _{\ell }^{\pm }\approx \omega _{\ell }\approx \omega
_{\ell 0}$, we obtain from Eq. (\ref{Eq15}) the result
\begin{equation}
\gamma _{\ell }(\omega _{\ell 0})\approx \frac{1}{2}V_{\ell }^{2}\left(
\omega _{\ell 0}\right) \sigma _{\ell }^{2}\left( \omega _{\ell 0}\right) =%
\frac{\Gamma _{\ell }}{2},  \label{Eq16}
\end{equation}%
while in the strong coupling regime, where $\omega _{\ell }^{+}\gg $ $\omega
_{\ell }^{-}$, we have
\begin{mathletters}
\begin{eqnarray}
\gamma _{\ell }(\omega _{\ell }^{+}) &\approx &\frac{1}{2}V_{\ell
}^{2}\left( \omega _{\ell }^{+}\right) \sigma _{\ell }^{2}\left( \omega
_{\ell }^{+}\right) ,  \label{Eq17a} \\
\gamma _{\ell }(\omega _{\ell }^{-}) &\approx &\frac{1}{4}V_{\ell
}^{2}\left( \omega _{\ell }^{-}\right) \sigma _{\ell }^{2}\left( \omega
_{\ell }^{-}\right) {\rm {\quad }}{\rm {when\;}}F=0  \label{Eq17b} \\
\gamma _{\ell }(\omega _{\ell }^{-}) &\approx &\frac{1}{2}V_{\ell
}^{2}\left( \omega _{\ell }^{-}\right) \sigma _{\ell }^{2}\left( \omega
_{\ell }^{-}\right) {\rm {\quad }}{\rm {when\;}}F\neq 0{\rm {.}}
\label{Eq17c}
\end{eqnarray}

>From the above results for the correlation functions, we observe that the
master equation for the strong coupling regime includes that for the weak
coupling regime. In fact, with $\gamma _{\ell }(\omega _{\ell }^{\pm
})\approx \Gamma _{\ell }/2$ and $N_{\ell }(\omega _{\ell }^{\pm })\approx
N_{\ell }(\omega _{\ell 0})$ we get the master equation for the weak
coupling regime from that for the strong coupling regime which, described in
the Schr\"{o}dinger picture via the mode operators $a_{\ell }$ and $a_{\ell
}^{\dagger }$, reads
\end{mathletters}
\begin{eqnarray}
\frac{d\rho _{12}}{dt} &=&\sum_{\ell }\left\{ i\left[ \rho _{12},\Omega
a_{\ell }^{\dagger }a_{\ell }+\lambda \sum_{m\neq \ell }a_{\ell }^{\dagger
}a_{m}+F\left( a_{2}^{\dagger }+a_{2}\right) \delta _{\ell 2}\right] \right.
\nonumber \\
&&+\frac{F}{2\left( \Omega ^{2}-\lambda ^{2}\right) }\left\{ \left( \Omega
\delta _{\ell 1}-\lambda \delta _{\ell 2}\right) \left[ \gamma _{\ell
}\left( \omega _{\ell }^{-}\right) -\gamma _{\ell }\left( \omega _{\ell
}^{+}\right) \right] +\left( \Omega \delta _{\ell 2}-\lambda \delta _{\ell
1}\right) \right.  \nonumber \\
&&\times \left. \left[ 2\gamma _{\ell }\left( \omega \right) -\gamma _{\ell
}\left( \omega _{\ell }^{-}\right) -\gamma _{\ell }\left( \omega _{\ell
}^{+}\right) \right] \right\} \left[ \rho _{12},a_{\ell }-a_{\ell }^{\dagger
}\right]  \nonumber \\
&&+\frac{1}{2}\left[ \gamma _{\ell }\left( \omega _{\ell }^{+}\right)
N_{\ell }\left( \omega _{\ell }^{+}\right) +\gamma _{\ell }\left( \omega
_{\ell }^{-}\right) N_{\ell }\left( \omega _{\ell }^{-}\right) \right]
\left( \left[ \left[ a_{\ell }^{\dagger },\rho _{12}\right] ,a_{\ell }\right]
+\left[ a_{\ell }^{\dagger },\left[ \rho _{12},a_{\ell }\right] \right]
\right)  \nonumber \\
&&+\frac{1}{2}\left[ \gamma _{\ell }\left( \omega _{\ell }^{+}\right)
+\gamma _{\ell }\left( \omega _{\ell }^{-}\right) \right] \left( \left[
a_{\ell }\rho _{12},a_{\ell }^{\dagger }\right] +\left[ a_{\ell },\rho
_{12}a_{\ell }^{\dagger }\right] \right)  \nonumber \\
&&+\frac{1}{2}\sum_{m\neq \ell }\left[ \gamma _{\ell }\left( \omega _{\ell
}^{+}\right) -\gamma _{\ell }\left( \omega _{\ell }^{-}\right) \right]
\left( \left[ a_{m}\rho _{12},a_{\ell }^{\dagger }\right] +\left[ a_{\ell
},\rho _{12}a_{m}^{\dagger }\right] \right)  \nonumber \\
&&\left. +\frac{1}{2}\left[ \gamma _{\ell }\left( \omega _{\ell }^{+}\right)
N_{\ell }\left( \omega _{\ell }^{+}\right) -\gamma _{\ell }\left( \omega
_{\ell }^{-}\right) N_{\ell }\left( \omega _{\ell }^{-}\right) \right]
\left( \left[ \left[ a_{2}^{\dagger },\rho _{12}\right] ,a_{1}\right] +\left[
a_{1}^{\dagger },\left[ \rho _{12},a_{2}\right] \right] \right) \right\} .
\label{Eq18}
\end{eqnarray}%
We stress that when the driving field is switched off, the second term in
Eq. (\ref{Eq18}) under the summation on $\ell $ disappears. Otherwise,
noting that $\Omega ^{2}-\lambda ^{2}=(\omega _{\ell }^{-}-\omega )(\omega
_{\ell }^{+}-\omega )$, when $F\neq 0$ and $\omega =\omega _{\ell }^{\pm }$,
this term becomes
\begin{equation}
\frac{F}{4\lambda }\left[ \gamma _{\ell }\left( \omega _{\ell }^{-}\right)
-\gamma _{\ell }\left( \omega _{\ell }^{+}\right) \right] \left( \delta
_{\ell 2}\mp \delta _{\ell 1}\right) \left[ \rho _{12},a_{\ell }-a_{\ell
}^{\dagger }\right] {\rm {.}}  \label{Eq18l}
\end{equation}

\qquad \qquad Next, we derive the master equations in the weak and strong
coupling regime. For simplicity, we define for the strong coupling regime $%
\gamma _{\ell }(\omega _{\ell }^{\pm })\equiv $ $\gamma _{\ell }^{\pm }$.
Besides, we assume from here on the resonance condition for the driving
field $\omega =\omega _{20}$.

\subsection{Weak coupling regime}

In the weak coupling regime, where $\omega _{\ell }^{\pm }\approx \omega
_{\ell }\approx \omega _{\ell 0}$, the resonant condition for the driving
field implies $\gamma _{\ell }\left( \omega \right) \approx \gamma _{\ell
}^{\pm }\approx \Gamma _{\ell }/2$, so that the master equation becomes
\begin{equation}
\frac{d\rho _{12}}{dt}=\frac{i}{\hbar }\left[ \rho _{12},{\sf H}_{0}\right]
+\sum_{\ell }{\cal L}_{\ell }\rho _{12}{\rm {,}}  \label{Eq19}
\end{equation}%
where
\begin{equation}
{\sf H}_{0}=\hbar \sum_{\ell }\left[ \Omega a_{\ell }^{\dagger }a_{\ell
}+\lambda \sum_{m\neq \ell }a_{\ell }^{\dagger }a_{m}+F\left( a_{2}^{\dagger
}+a_{2}\right) \delta _{\ell 2}\right] {\rm {,}}  \label{Eq20}
\end{equation}%
and the Liouville operator ${\cal L}_{\ell }\rho _{12}$ is given by the
usual operator structure
\begin{eqnarray}
{\cal L}_{\ell }\rho _{12} &\equiv &\frac{1}{2}\Gamma _{\ell }\left\{
N_{\ell }\left( \omega _{\ell 0}\right) \left( \left[ a_{\ell }^{\dagger
}\rho _{12},a_{\ell }\right] +\left[ a_{\ell }^{\dagger },\rho _{12}a_{\ell }%
\right] \right) \right.  \nonumber \\
&&+\left. \left( N_{\ell }\left( \omega _{\ell 0}\right) +1\right) \left(
\left[ a_{\ell }\rho _{12},a_{\ell }^{\dagger }\right] +\left[ a_{\ell
},\rho _{12}a_{\ell }^{\dagger }\right] \right) \right\} {\rm {.}}
\label{Eq21}
\end{eqnarray}%
Therefore, in the weak coupling regime and assuming that the driving field
is resonant with oscillator $2$, the interaction between the field modes and
the amplification process appears only in the von Neumann term of the master
equation, and does not affect the dissipative mechanism of the individual
cavities. However, when the driving field is out of resonance with
oscillator $2$, a correction term is added to the Liouville operator ${\cal L%
}_{\ell }\rho _{12}$, given by
\begin{equation}
\frac{F}{\left( \Omega ^{2}-\lambda ^{2}\right) }\left( \Omega \delta _{\ell
2}-\lambda \delta _{\ell 1}\right) \left[ \gamma _{\ell }\left( \omega
\right) -\gamma _{\ell }\left( \omega _{\ell 0}\right) \right] \left[ \rho
_{12},a_{\ell }-a_{\ell }^{\dagger }\right] .  \label{Eq21l}
\end{equation}%
For a strong amplification process ($F/\omega _{20}\approx 1$), this
correction becomes $F\Omega /\left( \Omega ^{2}-\lambda ^{2}\right) \approx
1/2$ for mode $\ell =2$, being ignored for mode $\ell =1$, since $F\lambda
/\left( \Omega ^{2}-\lambda ^{2}\right) \ll 1$.

\subsection{Strong coupling regime}

Below we present the master equation for the strong coupling regime
considering the general situation where both driving field and reservoir
temperature are present. However, as discussed above, when switching off the
driving field we must consider reservoirs at absolute zero for our results
to be valid. In this regime, where $\lambda /\omega _{\ell }\approx 1$, the
master equation is written
\begin{equation}
\frac{d\rho _{12}}{dt}=\frac{i}{\hbar }\left[ \rho _{12},{\sf H}_{0}\right]
+\sum_{\ell }{\cal L}_{\ell }\rho _{12}+{\cal L}_{12}\rho _{12},
\label{Eq22}
\end{equation}%
where
\begin{eqnarray}
{\cal L}_{\ell }\rho _{12} &=&\frac{1}{2}\left( \gamma _{\ell }^{+}N_{\ell
}\left( \omega _{\ell }^{+}\right) +\gamma _{\ell }^{-}N_{\ell }\left(
\omega _{\ell }^{-}\right) \right) \left( \left[ \left[ a_{\ell }^{\dagger
},\rho _{12}\right] ,a_{\ell }\right] +\left[ a_{\ell }^{\dagger },\left[
\rho _{12},a_{\ell }\right] \right] \right)  \nonumber \\
&&+\frac{1}{2}\left( \gamma _{\ell }^{+}+\gamma _{\ell }^{-}\right) \left( %
\left[ a_{\ell }\rho _{12},a_{\ell }^{\dagger }\right] +\left[ a_{\ell
},\rho _{12}a_{\ell }^{\dagger }\right] \right)  \nonumber \\
&&+\frac{F}{2\left( \Omega ^{2}-\lambda ^{2}\right) }\left\{ \left( \Omega
\delta _{\ell 1}-\lambda \delta _{\ell 2}\right) \left( \gamma _{\ell
}^{-}-\gamma _{\ell }^{+}\right) +\left( \Omega \delta _{\ell 2}-\lambda
\delta _{\ell 1}\right) \right.  \nonumber \\
&&\times \left. \left[ 2\gamma _{\ell }\left( \omega _{20}\right) -\gamma
_{\ell }^{-}-\gamma _{\ell }^{+}\right] \right\} \left[ \rho _{12},a_{\ell
}-a_{\ell }^{\dagger }\right] ,  \label{Eq23}
\end{eqnarray}%
and a cross-decay channel is included via the Liouville operator
\begin{eqnarray}
{\cal L}_{12}\rho _{12} &=&\frac{1}{2}\sum_{\ell }\left\{ \sum_{m\neq \ell
}\left( \gamma _{\ell }^{+}-\gamma _{\ell }^{-}\right) \left( \left[
a_{m}\rho _{12},a_{\ell }^{\dagger }\right] +\left[ a_{\ell },\rho
_{12}a_{m}^{\dagger }\right] \right) \right.  \nonumber \\
&&\left. +\left[ \gamma _{\ell }^{+}N_{\ell }\left( \omega _{\ell
}^{+}\right) -\gamma _{\ell }^{-}N_{\ell }\left( \omega _{\ell }^{-}\right) %
\right] \left( \left[ \left[ a_{2}^{\dagger },\rho _{12}\right] ,a_{1}\right]
+\left[ a_{1}^{\dagger },\left[ \rho _{12},a_{2}\right] \right] \right)
\right\} .  \label{Eq24}
\end{eqnarray}%
Note that in the weak coupling regime, where $\gamma _{\ell }^{\pm }\approx
\Gamma _{\ell }/2$, we obtain Eq.(\ref{Eq19}) from Eq. (\ref{Eq22}).
Evidently, the Liouville operator accounting for the cross-decay channel, $%
{\cal L}_{12}\rho _{12}$, owing to the strong coupling between the
oscillators, can be of the same order of magnitude as the direct-decay
channels ${\cal L}_{1}\rho _{12}$ and ${\cal L}_{2}\rho _{12}$. In this
regime, in contrast to the weak coupling case, a strong driving field ($%
F/\omega _{20}\approx 1$) modifies the Liouville operator ${\cal L}_{\ell
}\rho _{12}$, such that $F\Omega /\left( \Omega ^{2}-\lambda ^{2}\right)
,F\lambda /\left( \Omega ^{2}-\lambda ^{2}\right) \approx 1$, independently
of the resonance consition. We observe that for both reservoirs at absolute
zero, the cross-decay channel is lost for the case where $\gamma _{\ell
}^{+}=\gamma _{\ell }^{-}$, which may occur, as discussed below, depending
on the spectral density of the reservoirs. In what follows it will become
clear that the cross-decay channel represented by the Liouville operator (%
\ref{Eq24}) leads to interesting results concerning the decoherence process
in strongly coupled oscillators.

\subsection{The split of the damping rate}

It is interesting to note that for the coupled dissipative oscillators the
damping rate $\gamma _{\ell }\left( \omega _{\ell 0}\right) =\Gamma _{\ell
}/2$ for mode $\ell $ splits into $\gamma _{\ell }^{+}$ and $\gamma _{\ell
}^{-}$. To illustrate this mechanism, we assume a Lorentzian coupling $%
V_{\ell }$ between the oscillators and their respective reservoirs, such
that the damping function $\Gamma _{\ell }\left( \chi \right) $, centered on
frequency $\chi _{0}$, is given by

\begin{equation}
\Gamma _{\ell }\left( \chi \right) =\sigma _{\ell }^{2}\frac{\Upsilon }{%
\left( \chi -\chi _{0}\right) ^{2}+\Upsilon ^{2}}{\rm {,}}  \label{Eq25}
\end{equation}%
with the parameter $\Upsilon $ accounting for the spectral sharpness around
the mode frequency.{\large \ }From the above expression and remembering,
from Eq. (\ref{Eq16}), that the frequency $\omega _{\ell }$ splits into two
shifted frequencies $\omega _{\ell }^{\pm }$, we obtain the double
Lorentzian function depicted in Fig. 2
\begin{eqnarray}
\Gamma _{\ell }\left( \chi \right) &=&\frac{\Upsilon \sigma _{\ell }^{2}}{2}%
\left( \frac{1}{\left( \chi -\omega _{\ell }^{+}\right) ^{2}+\Upsilon ^{2}}+%
\frac{1}{\left( \chi -\omega _{\ell }^{-}\right) ^{2}+\Upsilon ^{2}}\right)
\nonumber \\
&=&\gamma _{\ell }\left( \omega _{\ell }^{+}\right) +\gamma _{\ell }\left(
\omega _{\ell }^{-}\right) {\rm {,}}  \label{Eq26}
\end{eqnarray}%
with maxima on $\omega _{\ell }^{\pm }$. In fact, from master equation (\ref%
{Eq18}), we observe that in the weak coupling regime, when $\gamma _{\ell
}\left( \omega _{\ell }^{+}\right) +\gamma _{\ell }\left( \omega _{\ell
}^{-}\right) =\Gamma _{\ell }(\omega _{\ell 0})=$ $\Gamma _{\ell }$ (since $%
\gamma _{\ell }\left( \omega _{\ell }^{+}\right) =\gamma _{\ell }\left(
\omega _{\ell }^{-}\right) =\Gamma _{\ell }/2$) , we obtain the expected
Liouville form for two independent dissipative oscillators. From Eq. (\ref%
{Eq26}) it is immediately obvious that in the weak coupling regime, where $%
\omega _{\ell }^{\pm }\approx \omega _{\ell 0}$, the damping function
presents only one peak, shown by the dotted line in Fig. 2. In this regime,
the damping rate, assumed to be the maximum of a sharp-peaked damping
function, i.e., $\Gamma _{\ell }$ (for a small value of $\Upsilon $),
becomes twice the value designated for $\gamma _{\ell }\left( \omega _{\ell
}^{\pm }\right) $. As $\lambda =\pm (\omega _{\ell }^{\pm }-\omega _{\ell })$
increases, the damping function splits into two Lorentzian functions whose
peak heights are half the original value $\Gamma _{\ell }$, as dictated by
the master equation (\ref{Eq22}) and shown by the solid line in Fig. 2. The
dashed line shows the situation where the two peaks can be clearly
distinguished, on the way to the strong coupling regime, $\lambda /\omega
_{\ell 0}\approx 1$, where the peak centered on $\omega _{\ell }^{-}$ shifts
to around the value $F\left[ \omega _{10}/\left( \omega _{10}+\omega
_{20}\right) \right] $, which can be made as smaller as we wish by
decreasing the amplitude of the driving field. In practice, the effect of
the strong coupling between the oscillators is essentially to shift the
normal-mode frequency $\omega _{\ell }^{\pm }$ to regions far way from the
natural frequency of the oscillator $\omega _{\ell 0}$, where the spectral
density of the reservoir may be significantly different from that around $%
\omega _{\ell 0}$. In this connection, the spectral density of the reservoir
$\sigma _{\ell }\left( \omega _{\ell k}\right) $ plays a crucial role in the
dissipative dynamics of strongly coupled oscillators, since the magnitude of
\ the damping rate $\gamma _{\ell }\left( \omega _{\ell }\right) $ depends
on $\sigma _{\ell }\left( \omega _{\ell }\right) $. For this reason, we next
analyze reservoirs with different spectral densities in order to illustrate
the interesting features arising from the strong coupling regime. Evidently,
the physical systems under consideration and their dissipative mechanisms
(i.e., the nature of the reservoirs and their spectral densities), will be
decisive for our analysis.

\section{Spectral densities of the reservoirs}

It is possible that specific spectral densities could be achieved through
engineered reservoirs, a program which has recently attracted considerable
attention also in attempts to control the decoherence process of quantum
states \cite{Zoller,Wineland1,Matos}. Therefore, the results we present
below, depending crucially on the spectral density of the reservoir, might
provide a motivation for future theoretical proposals on engineered
reservoirs. For the following discussion we consider the strong coupling
regime and set both driving field and reservoir temperatures to zero ($F,T=0$%
) and remember, from the definition $\gamma _{\ell }(\omega _{\ell }^{\pm
})\equiv $ $\gamma _{\ell }^{\pm }$, that the parameter $\gamma _{\ell
}^{\pm }$\ depends on the reservoir spectral density around $\omega _{\ell
}^{\pm }$.

\subsection{Markovian white noise}

We start with the simplest case of Markovian white noise, where the spectral
density of the reservoir is invariant over translation in frequency space,
as depicted in Fig 3(a). In this case, assuming a Lorentzian coupling
between the oscillators and their respective reservoirs, centered around the
effective frequencies $\omega _{\ell }^{\pm }$, as in Eq. (\ref{Eq26}), we
get $\gamma _{\ell }^{-}=\gamma _{\ell }^{+}/2=\Gamma _{\ell }/4$. In fact,
the effective frequency $\omega _{\ell }^{-}$ shifts to around zero (when $%
F=0)$, and so, the system-reservoir coupling $\gamma _{\ell }^{-}$ becomes
half the value $\gamma _{\ell }^{+}=\Gamma _{\ell }/2$, as can be concluded
immediately from Eqs. (\ref{Eq17a}), (\ref{Eq17b}) and \ref{Eq17c}. The
system-reservoir couplings around $\omega _{\ell }^{\pm }$ are represented
by shaded regions in Fig. 3(a). Therefore, for a Markovian white noise
reservoir, strong coupling between the oscillators delays the decoherence
time of a joint state which is an eigenstate of normal mode $\omega _{\ell
}^{-}$, as will be discussed latter. Next, we analyze two cases of
non-Markovian colored noise, still assuming that $V_{\ell }$, $\sigma _{\ell
}$,\ and $N_{\ell }$\ are functions that vary slowly around the frequency $%
\omega _{\ell }^{\pm }$, as discussed above.

\subsection{A Lorentzian spectral density}

Let us consider a Lorentzian spectral density of the reservoir which goes to
zero at both effective frequencies $\omega _{\ell }^{-}$ and $\omega _{\ell
}^{+}$ (Fig. 3(b)). The achievement of a reservoir spectral density for
which $\sigma _{\ell }\left( \omega _{\ell }^{\pm }\right) \approx 0$ would
result in a damping function $\gamma _{\ell }(\omega _{\ell }^{\pm })$
arising from both terms on the right hand side of Eq. (\ref{Eq26}) (i.e., $%
\Upsilon \sigma _{\ell }^{2}/\left[ \left( \chi -\omega _{\ell }^{\pm
}\right) ^{2}+\Upsilon ^{2}\right] \approx 0$), neither of which would
contribute significantly to the relaxation process. Therefore, for such a
Lorentzian coupling between the field modes and their respective reservoirs,
we get from Eqs. (\ref{Eq17a}), (\ref{Eq17b}) and \ref{Eq17c} the rates $%
\gamma _{\ell }^{\pm }\ll \Gamma _{\ell }$, since both effective frequencies
$\omega _{\ell }^{+}$ and $\omega _{\ell }^{-}$ shift to regions where the
spectral densities of the reservoir are considerable smaller than that
around $\omega _{\ell 0}$. In view of the difficulty of obtaining a
Lorentzian spectral density sufficiently small around $\omega _{\ell }^{+}$,
such that $\sigma _{\ell }\left( \omega _{\ell }^{+}\right) \approx 0$, we
next turn to a more realistic case, which we call a wide Lorentzian spectral
density.

\subsection{A wide Lorentzian spectral density}

Finally, we consider a wide Lorentzian spectral density, which goes to zero
at the effective frequency $\omega _{\ell }^{-}$, reaches a flat maximum
around $\omega _{\ell 0}$, continues on it up to around $\omega _{\ell }^{+}$
and only reaches zero again at frequencies far beyond $\omega _{\ell }^{+}$,
as depicted in Fig. 3(c). Differently from the Lorentzian spectral density,
in this case only the damping function $\gamma _{\ell }^{-}$ arising from
the second term on the right hand side of Eq. (\ref{Eq26}) (i.e., $\Upsilon
\sigma _{\ell }^{2}/\left[ \left( \chi -\omega _{\ell }^{-}\right)
^{2}+\Upsilon ^{2}\right] \approx 0$) does not contribute significantly to
the relaxation process; i.e., $\gamma _{\ell }^{-}\ll \Gamma _{\ell }$,
since the resonances around the effective frequency $\omega _{\ell }^{-}$
can be disregarded. Only the system-reservoir coupling around the effective
frequency $\omega _{\ell }^{+}$ contributes to the relaxation process of the
cavity mode ($\gamma _{\ell }^{+}=\Gamma _{\ell }/2$). Such a model can be
considered as similar to subhomic dissipation \cite{CL,Leggett} and can be
reasonably applied for a variety of phonon-like spectral densities where the
de Debye model applies very well in the domain of small frequencies.

\section{The Fokker-Planck equation}

Using the standard procedures, we derive a c-number version of the master
equation (\ref{Eq18}) for the Glauber P-function \cite{MLW,WM} given by
\begin{equation}
\frac{dP}{dt}=\frac{1}{2}\sum_{\ell ,m}\left[ \Pi _{\ell }+C_{\ell }(\alpha
_{1},\alpha _{2})\frac{\partial }{\partial \alpha _{\ell }}+D_{\ell m}\frac{%
\partial ^{2}}{\partial \alpha _{\ell }\partial \alpha _{m}^{\ast }}+H.c.%
\right] P{\rm {,}}  \label{Eq27}
\end{equation}%
where the function $C_{\ell }(\alpha _{1},\alpha _{2})$ and the matrix
elements $D_{\ell m}$ satisfy
\begin{mathletters}
\begin{eqnarray}
C_{\ell }(\alpha _{1},\alpha _{2}) &=&B_{\ell }+\alpha _{\ell }E_{\ell
}^{+}+\alpha _{\ell -(-1)^{\ell }}E_{\ell }^{-},  \label{Eq28a} \\
D_{\ell \ell } &=&\gamma _{\ell }^{+}N_{\ell }\left( \omega _{\ell
}^{+}\right) +\gamma _{\ell }^{-}N_{\ell }\left( \omega _{\ell }^{-}\right) ,
\label{Eq28b} \\
D_{12} &=&D_{21}=\frac{1}{2}\sum_{\ell }\left[ \gamma _{\ell }^{+}N_{\ell
}\left( \omega _{\ell }^{+}\right) -\gamma _{\ell }^{-}N_{\ell }\left(
\omega _{\ell }^{-}\right) \right] ,  \label{Eq28c}
\end{eqnarray}%
while the parameters $\Pi _{\ell }$, $E_{\ell }^{\pm }$\ and $B_{\ell }$\
are defined by{\LARGE \ }
\end{mathletters}
\begin{eqnarray}
\Pi _{\ell } &=&\frac{1}{2}\left( \gamma _{\ell }^{-}+\gamma _{\ell
}^{+}\right) ,  \label{Eq29a} \\
E_{\ell }^{+} &=&\frac{1}{2}\left( \gamma _{\ell }^{+}+\gamma _{\ell
}^{-}+2i\Omega \right) ,  \label{Eq29b} \\
E_{\ell }^{-} &=&\frac{1}{2}\left( \gamma _{\ell }^{+}-\gamma _{\ell
}^{-}+2i\lambda \right) ,  \label{Eq29c} \\
B_{\ell } &=&iF\delta _{\ell 2}-\frac{F}{2\left( \Omega ^{2}-\lambda
^{2}\right) }\left[ \left( \Omega \delta _{\ell 2}-\lambda \delta _{\ell
1}\right) \left( 2\gamma _{\ell }(\omega _{20})-\gamma _{\ell }^{-}-\gamma
_{\ell }^{+}\right) \right.  \nonumber \\
&&+\left. \left( \Omega \delta _{\ell 1}-\lambda \delta _{\ell 2}\right)
\left( \gamma _{\ell }^{-}-\gamma _{\ell }^{+}\right) \right] {.}
\label{Eq29d}
\end{eqnarray}%
Note that the drift coefficient $C_{\ell }$ is linearly dependent upon both
variables $\alpha _{1}$ and $\alpha _{2}$, while the diffusion coefficient $%
D_{mn}$ is constant, determining an Orstein-Uhlenbeck process \cite{Gardiner}%
.

In an extreme case of the weak coupling regime, which we do not consider in
the present work, where $\lambda \lesssim $ $\Gamma _{\ell }$, the drift
coefficient $C_{\ell }$ depends mainly on $\alpha _{\ell }$ and the
diffusion coefficient is given by $D_{mn}=\delta _{mn}D_{mm}$, which is
strictly positive. In this particular case, the dynamics of the two
oscillators decouple, since the two-mode state relaxes, due to the damping
process, before they have time to interact. Thus
\begin{equation}
\frac{dP}{dt}=\sum_{\ell }\left[ \Pi _{\ell }+C_{\ell }(\alpha _{\ell })%
\frac{\partial }{\partial \alpha _{\ell }}+\frac{D_{\ell \ell }}{2}\frac{%
\partial ^{2}}{\partial \alpha _{\ell }\partial \alpha _{\ell }^{\ast }}+H.c.%
\right] P,  \label{Eq30}
\end{equation}%
and the general solution is simply the summation of the independent
solutions for each mode.

\subsection{Solution of the Fokker-Planck equation at absolute zero}

>From now on we assume both reservoirs set to absolute zero, such that $%
N_{\ell }(\omega _{\ell }^{\pm })=0$. This assumption is equivalent to
setting $D_{mn}=0$, and thus the Fokker-Planck equation (\ref{Eq27}) reduces
to a simple drift equation
\begin{equation}
\frac{dP}{dt}=\sum_{\ell }\left[ \Pi _{\ell }+C_{\ell }(\alpha _{1},\alpha
_{2})\frac{\partial }{\partial \alpha _{\ell }}+H.c.\right] P.  \label{Eq31}
\end{equation}

With the substitution $P=P^{\prime }\exp \left[ 2\left( \Pi _{1}+\Pi
_{2}\right) t\right] $, we simplify the above equation to the form
\begin{equation}
\frac{dP^{\prime }}{dt}=\sum_{\ell }\left[ C_{\ell }(\alpha _{1},\alpha _{2})%
\frac{\partial }{\partial \alpha _{\ell }}+H.c.\right] P^{\prime }{\rm {,}}
\label{Eq32}
\end{equation}%
and assuming that $P^{\prime }(\alpha _{1},\alpha _{2},t)=P^{\prime }(\alpha
_{1}(t),\alpha _{2}(t))$,{\Large \ }we get
\begin{equation}
\frac{d}{dt}P^{\prime }(\alpha _{1}(t),\alpha _{2}(t))=\sum_{\ell }\left(
\frac{\partial \alpha _{\ell }}{\partial t}\frac{\partial }{\partial \alpha
_{\ell }}+H.c\right) P^{\prime }(\alpha _{1}(t),\alpha _{2}(t)){\rm {.}}
\label{Eq33}
\end{equation}%
Therefore, from Eqs. (\ref{Eq32}) and (\ref{Eq33}) we obtain the system of
coupled equations
\begin{mathletters}
\begin{eqnarray}
\frac{\partial \alpha _{1}}{\partial t} &=&B_{1}+\alpha _{1}E_{1}^{+}+\alpha
_{2}E_{1}^{-},  \label{Eq34a} \\
\frac{\partial \alpha _{2}}{\partial t} &=&B_{2}+\alpha _{2}E_{2}^{+}+\alpha
_{1}E_{2}^{-},  \label{Eq34b}
\end{eqnarray}%
which leads to
\end{mathletters}
\begin{equation}
\frac{\partial ^{2}\alpha _{1}}{\partial t^{2}}-\left(
E_{1}^{+}+E_{2}^{+}\right) \frac{\partial \alpha _{1}}{\partial t}+\alpha
_{1}\left( E_{1}^{+}E_{2}^{+}-E_{1}^{-}E_{2}^{-}\right) +\left(
B_{1}E_{2}^{+}-B_{2}E_{1}^{-}\right) =0.  \label{Eq34l}
\end{equation}

Next, we define the parameters (where we have disregarded terms of order O$%
^{2}$($\gamma _{\ell }^{\pm }/\lambda $)){\LARGE \ }$\Lambda =\left( \gamma
_{1}^{+}+\gamma _{2}^{+}+\gamma _{1}^{-}+\gamma _{2}^{-}\right) /4+i\Omega $%
, $\Delta =\left( \gamma _{1}^{+}-\gamma _{2}^{+}+\gamma _{1}^{-}-\gamma
_{2}^{-}\right) /4$, $\Phi =\left( \gamma _{1}^{+}+\gamma _{2}^{+}-\gamma
_{1}^{-}-\gamma _{2}^{-}\right) /4$, $\Theta =\left( \gamma _{1}^{+}-\gamma
_{2}^{+}-\gamma _{1}^{-}+\gamma _{2}^{-}\right) /4$, and the time-dependent
functions
\begin{mathletters}
\begin{eqnarray}
W^{\pm } &=&\cosh \Phi t\left( \cos \lambda t\pm \frac{\Delta }{\lambda }%
\sin \lambda t\right) +i\sinh \Phi t\left( \sin \lambda t\mp \frac{\Delta }{%
\lambda }\cos \lambda t\right) ,  \label{Eq35a} \\
Z^{\pm } &=&\left( \sinh \Phi t\cos \lambda t\pm \frac{\Theta }{\lambda }%
\cosh \Phi t\sin \lambda t\right) +i\left( \cosh \Phi t\sin \lambda t\mp
\frac{\Theta }{\lambda }\sinh \Phi t\cos \lambda t\right) .  \label{Eq35b}
\end{eqnarray}%
Defining also the parameters
\end{mathletters}
\begin{eqnarray}
G_{1} &=&\frac{B_{2}E_{1}^{-}-B_{1}E_{2}^{+}}{%
E_{1}^{+}E_{2}^{+}-E_{1}^{-}E_{2}^{-}},  \label{Eq36a} \\
G_{2} &=&\frac{B_{1}E_{2}^{-}-B_{2}E_{1}^{+}}{%
E_{1}^{+}E_{2}^{+}-E_{1}^{-}E_{2}^{-}},  \label{Eq36b}
\end{eqnarray}%
we finally obtain the solution of Eqs. (\ref{Eq34a}, \ref{Eq34b}), written
in the compact form
\begin{mathletters}
\begin{eqnarray}
\alpha _{1}(t) &=&%
\mathop{\rm e}%
{}^{\Lambda t}\left[ \left( \alpha _{1}^{0}-G_{1}\right) W^{+}+\left( \alpha
_{2}^{0}-G_{2}\right) Z^{+}\right] +G_{1}{\rm {,}}  \label{Eq37a} \\
\alpha _{2}(t) &=&%
\mathop{\rm e}%
{}^{\Lambda t}\left[ \left( \alpha _{2}^{0}-G_{2}\right) W^{-}+\left( \alpha
_{1}^{0}-G_{1}\right) Z^{-}\right] +G_{2}{\rm {,}}  \label{Eq37b}
\end{eqnarray}%
where $\alpha _{\ell }^{0}$ indicates $\alpha _{\ell }(t=0)$. From the
solution of Eqs. (\ref{Eq37a}) and (\ref{Eq37b}) it follows that the
Fokker-Planck equation (\ref{Eq31}) can be solved to give
\end{mathletters}
\begin{equation}
P(\alpha _{1},\alpha _{2},t)=%
\mathop{\rm e}%
{}^{2\left( \Pi _{1}+\Pi _{2}\right) t}P(\alpha _{1},\alpha _{2},0)|_{\alpha
_{\ell }\rightarrow \alpha _{\ell }(t)}{\rm {,}}  \label{Eq38}
\end{equation}%
where $P(\alpha _{1},\alpha _{2},0)$ is the P-function for the initial
state. The evolved P-function is then obtained by simply replacing $\alpha
_{\ell }$ by the evolved parameters given by Eqs. (\ref{Eq37a}) and (\ref%
{Eq37b}). Finally, from Eq. (\ref{Eq38}) the evolved density operator
related to (\ref{Eq18}) follows from
\begin{equation}
\rho _{12}(t)=\int d^{2}\alpha _{1}d^{2}\alpha _{2}P(\alpha _{1},\alpha
_{2},t)|\alpha _{1},\alpha _{2}\rangle \langle \alpha _{1},\alpha _{2}|.
\label{Eq39}
\end{equation}

\subsection{Initial joint states}

Next, we obtain the density operator $\rho _{12}(t)$ supposing that the
modes $1$ and $2$ are prepared in a superposition of coherent states of the
form
\begin{equation}
\left| \varphi \right\rangle ={\cal N}_{\pm }\left( \left| \beta
_{I}^{1},\beta _{I}^{2}\right\rangle \pm \left| \beta _{II}^{1},\beta
_{II}^{2}\right\rangle \right) ,  \label{Eq40}
\end{equation}%
where ${\cal N}_{\pm }$ stands for the normalization factor and the
subscripts $I,II$ are related to different coherent states participating in
the superposition. From Eqs. (\ref{Eq39}) and (\ref{Eq40}), the
corresponding evolved density operator is given by
\begin{equation}
\rho _{12}(t)=\sum_{m,n=I,II}C_{mn}(t)\left| \varsigma _{m},\zeta
_{m}\right\rangle \left\langle \varsigma _{n},\zeta _{n}\right| ,
\label{Eq41}
\end{equation}%
where the labels $\varsigma $ and $\zeta $ represent the states of modes $1$
and $2$, respectively. The expansion coefficients read
\begin{eqnarray}
C_{mn}(t) &=&{\cal N}_{\pm }^{2}\left( \pm 1\right) ^{1-\delta
_{mn}}\left\langle \beta _{n}^{1}\right. \left| \beta _{m}^{1}\right\rangle
^{1-\exp \left( -2{\rm Re}\left\{ \Lambda \right\} t\right) \left( \left|
W^{-}\right| ^{2}+\left| Z^{-}\right| ^{2}\right) }  \nonumber \\
&&\times \left\langle \beta _{n}^{2}\right. \left| \beta
_{m}^{2}\right\rangle ^{1-\exp \left( -2{\rm Re}\left\{ \Lambda \right\}
t\right) \left( \left| W^{+}\right| ^{2}+\left| Z^{+}\right| ^{2}\right) }%
\mathop{\rm e}%
{}^{i(\theta _{mn}^{(1)}+\theta _{mn}^{(3)})+\theta _{mn}^{(2)}+\theta
_{mn}^{(4)}},  \label{Eq42}
\end{eqnarray}%
while the states $\varsigma $ and $\zeta $ are given by
\begin{mathletters}
\begin{eqnarray}
\varsigma _{\ell } &=&%
\mathop{\rm e}%
{}^{-\Lambda t}\left[ \left( \beta _{\ell }^{1}-G_{1}\right) W^{-}-\left(
\beta _{\ell }^{2}-G_{2}\right) Z^{+}\right] +G_{1},  \label{Eq43a} \\
\zeta _{\ell } &=&%
\mathop{\rm e}%
{}^{-\Lambda t}\left[ \left( \beta _{\ell }^{2}-G_{2}\right) W^{+}-\left(
\beta _{\ell }^{1}-G_{1}\right) Z^{-}\right] +G_{2}  \label{Eq43b}
\end{eqnarray}
We define the time-dependent functions
\end{mathletters}
\begin{mathletters}
\begin{eqnarray}
\theta _{mn}^{(1)} &=&{\rm Im}\left\{
\mathop{\rm e}%
{}^{-\Lambda ^{\ast }t}\left[ \left( W^{+}\right) ^{\ast }\left( \beta
_{m}^{2\ast }-\beta _{n}^{2\ast }\right) -\left( Z^{-}\right) ^{\ast }\left(
\beta _{m}^{1\ast }-\beta _{n}^{1\ast }\right) \right] \right.  \nonumber \\
&&\times \left. \left[
\mathop{\rm e}%
{}^{-\Lambda t}\left( G_{1}Z^{-}-G_{2}W^{+}\right) +G_{2}\right] \right\} ,
\label{Eq44a} \\
2\theta _{mn}^{(2)} &=&%
\mathop{\rm e}%
{}^{-2{\rm Re}\left\{ \Lambda \right\} t}\left\{ W^{+}\left( Z^{-}\right)
^{\ast }\left[ \beta _{n}^{1\ast }\left( \beta _{m}^{2}-\beta
_{n}^{2}\right) -\left( \beta _{m}^{1\ast }-\beta _{n}^{1\ast }\right) \beta
_{m}^{2}\right] \right.  \nonumber \\
&&-\left. \left( W^{+}\right) ^{\ast }Z^{-}\left[ \beta _{m}^{1}\left( \beta
_{m}^{2\ast }-\beta _{n}^{2\ast }\right) -\left( \beta _{m}^{1}-\beta
_{n}^{1}\right) \beta _{n}^{2\ast }\right] \right\} ,  \label{Eq44b} \\
\theta _{mn}^{(3)} &=&{\rm Im}\left\{
\mathop{\rm e}%
{}^{-\Lambda ^{\ast }t}\left[ \left( W^{-}\right) ^{\ast }\left( \beta
_{m}^{1\ast }-\beta _{n}^{1\ast }\right) -\left( Z^{+}\right) ^{\ast }\left(
\beta _{m}^{2\ast }-\beta _{n}^{2\ast }\right) \right] \right.  \nonumber \\
&&\times \left. \left[
\mathop{\rm e}%
{}^{-\Lambda t}\left( G_{2}Z^{+}-G_{1}W^{-}\right) +G_{1}\right] \right\} ,
\label{Eq44c} \\
2\theta _{mn}^{(4)} &=&%
\mathop{\rm e}%
{}^{-2{\rm Re}\left\{ \Lambda \right\} t}\left\{ W^{-}\left( Z^{+}\right)
^{\ast }\left[ \left( \beta _{m}^{1}-\beta _{n}^{1}\right) \beta _{n}^{2\ast
}-\beta _{m}^{1}\left( \beta _{m}^{2\ast }-\beta _{n}^{2\ast }\right) \right]
\right.  \nonumber \\
&&-\left. \left( W^{-}\right) ^{\ast }Z^{+}\left[ \left( \beta _{m}^{1\ast
}-\beta _{n}^{1\ast }\right) \beta _{m}^{2}-\beta _{n}^{1\ast }\left( \beta
_{m}^{2}-\beta _{n}^{2}\right) \right] \right\} .  \label{Eq44d}
\end{eqnarray}%
Finally, the reduced density operator can be obtained easily from Eq. (\ref%
{Eq41}), being
\end{mathletters}
\begin{eqnarray}
\rho _{1}(t) &=&{}{\cal N}_{\pm }^{2}\sum_{m,n=I,II}\left( \pm 1\right)
^{1-\delta _{mn}}\left\langle \beta _{n}^{1}\right. \left| \beta
_{m}^{1}\right\rangle ^{1-\exp \left( -2{\rm Re}\left\{ \Lambda \right\}
t\right) \left| W^{-}\right| ^{2}}  \nonumber \\
&&\times \left\langle \beta _{n}^{2}\right| \left. \beta
_{m}^{2}\right\rangle ^{1-\exp \left( -2{\rm Re}\left\{ \Lambda \right\}
t\right) \left| Z^{+}\right| ^{2}}%
\mathop{\rm e}%
{}^{i\theta _{mn}^{(3)}+\theta _{mn}^{(4)}}\left| \varsigma
_{m}\right\rangle \left\langle \varsigma _{n}\right| {\rm {,}}  \label{Eq45a}
\\
\rho _{2}(t) &=&{\cal N}_{\pm }^{2}\sum_{m,n=I,II}\left( \pm 1\right)
^{1-\delta _{mn}}\left\langle \beta _{n}^{1}\right| \left. \beta
_{m}^{1}\right\rangle ^{1-\exp \left( -2{\rm Re}\left\{ \Lambda \right\}
t\right) \left| Z^{-}\right| ^{2}}  \nonumber \\
&&\times \left\langle \beta _{n}^{2}\right. \left| \beta
_{m}^{2}\right\rangle ^{1-\exp \left( -2{\rm Re}\left\{ \Lambda \right\}
t\right) \left| W^{+}\right| ^{2}}%
\mathop{\rm e}%
{}^{i\theta _{mn}^{(1)}+\theta _{mn}^{(2)}}\left| \zeta _{m}\right\rangle
\left\langle \zeta _{n}\right| .  \label{Eq45b}
\end{eqnarray}

\section{State recurrence and swap dynamics}

Let us suppose that the joint system $1+2$ is prepared in the superposition
state
\begin{equation}
|\psi _{12}\rangle ={}{\cal N}_{\pm }\left( |\alpha \rangle \pm |-\alpha
\rangle \right) _{1}\otimes |\eta \rangle _{2}{\rm {,}}  \label{Eq46}
\end{equation}
which represents a particular case of Eq. (\ref{Eq40}), where $\beta
_{I}^{1}=-\beta _{II}^{1}=\alpha $ and $\beta _{I}^{2}=\beta _{II}^{2}=\eta
. $

In this section we analyze, in the weak and strong coupling regimes,\ the
effect of dissipation on two phenomena: $i)$ the joint-system state
recurrence and $ii)$ the state swap between the systems. Considering the
dynamics of the coupled systems, governed by the Fokker-Planck equation (\ref%
{Eq31}), we calculate $i)$ the probability that in each system the initial
state recurs -- the recurrence time for both systems being the same -- and $%
ii)$ the probability of a swapping of states between the systems, i.e., the
probability\ of oscillator $1$ being in the initial state of oscillator $2$,
and vice-versa. Recurrence is guaranteed whenever a dynamical system is
closed and its evolution is limited \cite{Bocchieri}, while state swap is
only possible for specific interactions. Here the two oscillators are
coupled by a bilinear Hamiltonian which, in principle, allows state swap to
occur between the systems \cite{meu}.

Considering the initial superposition (\ref{Eq46}), the time-evolved joint
state following from the density operator (\ref{Eq41}) reads
\begin{equation}
\rho _{12}(t)=\sum_{m,n=I,II}{\cal C}_{mn}(t)\left| \varsigma _{m},\zeta
_{m}\right\rangle \left\langle \varsigma _{n},\zeta _{n}\right| {\rm {,}}
\label{Eq47}
\end{equation}%
where
\begin{eqnarray}
{\cal C}_{mn}(t) &=&{\cal N}_{\pm }^{2}\left( \pm 1\right) ^{1-\delta
_{mn}}\exp \left\{ -2\left| \alpha \right| ^{2}\left[ 1-\left( \left|
W^{-}\right| ^{2}+\left| Z^{-}\right| ^{2}\right) \exp \left( -2{\rm Re}%
\left\{ \Lambda \right\} t\right) \right] \right\}   \nonumber \\
&&\times \exp \left\{ i\left[ \theta _{mn}^{(1)}+\theta _{mn}^{(3)}+%
\mathop{\rm Im}%
\left( \theta _{mn}^{(2)}+\theta _{mn}^{(4)}\right) \right] \right\} {}{\rm {%
,}}  \label{Eq48}
\end{eqnarray}%
and the probability of recurrence is given by
\begin{eqnarray}
P_{R}(t) &\equiv &Tr_{12}\left[ \rho _{12}(t)\rho _{12}(0)\right]   \nonumber
\\
&=&{\cal N}_{\pm }^{2}\sum_{m,n=I,II}{\cal C}_{mn}(t)\left\langle \varsigma
_{n}\right| \left( |\alpha \rangle \pm |-\alpha \rangle \right) _{1}\left(
\left\langle \alpha \right| \pm \left\langle -\alpha \right| \right) \left|
\varsigma _{m}\right\rangle \left\langle \zeta _{m}\right| \left. \eta
\right\rangle _{2}\left\langle \eta \right| \left. \zeta _{m}\right\rangle
{\rm {.}}  \label{Eq49}
\end{eqnarray}%
Analogously to the recurrence probability, the state-swap probability is
calculated by swapping the labels of each initial state of the two fields,
as defined in \cite{meu}:
\begin{eqnarray}
P_{S}(t) &\equiv &Tr\left[ \rho _{12}(t)\rho _{12}(0)|_{1(2)\rightarrow 2(1)}%
\right]   \nonumber \\
&=&{\cal N}_{\pm }^{2}\sum_{m,n=I,II}{\cal C}_{mn}(t)\left\langle \varsigma
_{n}\right| \left. \eta \right\rangle _{2}\left\langle \eta \right. \left|
\varsigma _{m}\right\rangle \left\langle \zeta _{m}\right| \left( |\alpha
\rangle \pm |-\alpha \rangle \right) _{1}\left( \left\langle \alpha \right|
\pm \left\langle -\alpha \right| \right) \left| \zeta _{m}\right\rangle {\rm
{.}}  \label{Eq50}
\end{eqnarray}

In Fig. 4 we plot the state-swap probability $P_{S}(t)$ (dotted line) and
the recurrence probability $P_{R}(t)$ (solid line) against the scaled time $%
\lambda t$, taking $\alpha =\eta =1$ as real parameters. In Figs. 3(a, b and
c) we set the driving field strength $F$ to zero (so that $\omega
_{10}=\omega _{20}$ and the minimum of $\omega _{\ell }^{-}$, occurring for $%
\lambda =2\omega _{20}$, also becomes zero) and assume the absence of
dissipation, so as to take these figures as references. In Fig. 4(a) we
consider the weak coupling regime, assuming $\lambda /\omega _{10}=2\times
10^{-2}$ (a somewhat exaggerated ratio used to visualize better the strong
oscillations of probabilities $P_{R}(t)$ and $P_{S}(t)$). We observe (dotted
line), as can be deduced from Eq. (\ref{Eq50}), that the modes will swap
their states whenever
\begin{equation}
\lambda t_{S}=(2n+1)\frac{\pi }{2},{\rm {\ }}n=0,1,2,...,  \label{Eq51}
\end{equation}%
which means that in the swap time $t_{S}$, the state of mode $1$ becomes $%
|\beta \rangle _{1}$, while the state of mode $2$ becomes the superposition $%
\left( |\alpha \rangle \pm |-\alpha \rangle \right) _{2}$. From the solid
line and Eq. (\ref{Eq49}), we observe that the joint-system state recurs
whenever
\begin{equation}
\lambda t_{R}=n\pi ,{\rm {\ }}n=0,1,2,...,  \label{Eq52}
\end{equation}%
i.e., in the recurrence time $t_{R}$ the mode $1$($2$) returns to its
initial state $\left( |\alpha \rangle \pm |-\alpha \rangle \right) _{1}$ ($%
|\beta \rangle _{2}$). It is evident that the shape of Fig. 4(a) results
mainly from the small value of $\lambda $\ (compared to $\Omega $) which
defines an envelope function.

In Fig. 4(b) an intermediate coupling is assumed, such that $\lambda /\omega
_{10}=1$. As can be observed, the recurrence dynamics remains the same
(solid line), while the swap dynamics begins to be affected by the coupling
strength. In Fig. 4(c) we consider the strong coupling regime where $\lambda
=2\omega _{10}$ ($\omega _{10}=\omega _{20}$), such that $\Omega /\lambda =1$
(note that $\Omega /\lambda =\omega _{10}/\lambda +\lambda /4\omega _{20}$).
As in Fig. 4(a), in Fig. 4(c) the recurrence process remains unchanged,
still obeying Eq. (\ref{Eq52}), so that the recurrence time becomes smaller
due to the strong coupling parameter $\lambda $. However, the swap dynamics
is almost completely lost, the remaining oscillations arising from the
nonorthogonality between the states $\left( |\alpha \rangle \pm |-\alpha
\rangle \right) _{1}$ and $|\beta \rangle _{2}$.

To understand the behavior of $P_{R}(t)$ and $P_{S}(t)$ in Figs. 4(b and c),
we note that in the absence of dissipation the recurrence and state-swap
probabilities are computed as $P_{R}(t)=\left| \left\langle \psi
_{12}(0)\right. \left| \psi _{12}(t)\right\rangle \right| ^{2}$ and $%
P_{S}(t)=\left| \left\langle \psi _{12}(0)\right. \left| \left. \psi
_{12}(t)\right| _{1(2)\rightarrow 2(1)}\right\rangle \right| ^{2}$,
respectively. For $P_{R}(t)$, the probability amplitude
\begin{eqnarray}
\left\langle \psi _{12}(0)\right. \left| \psi _{12}(t)\right\rangle &=&{\cal %
N}_{\pm }^{2}\left( \left\langle \alpha \right. \left| \varsigma
_{I}\right\rangle \left\langle \eta \right. |\zeta _{I}\rangle +\exp \left(
2i\alpha G_{2}\sin \left( \lambda t\right) \right) \left\langle -\alpha
\right. \left| \varsigma _{II}\right\rangle \left\langle \eta \right. |\zeta
_{II}\rangle \right.  \nonumber \\
&&\left. \pm \left\langle -\alpha \right. \left| \varsigma _{I}\right\rangle
\left\langle \eta \right. |\zeta _{I}\rangle \pm \exp \left( 2i\alpha
G_{2}\sin \left( \lambda t\right) \right) \left\langle \alpha \right. \left|
\varsigma _{II}\right\rangle \left\langle \eta \right. |\zeta _{II}\rangle
\right) ,  \label{Eq53}
\end{eqnarray}%
tends to unity under the conditions $\zeta _{I},\zeta _{II}\rightarrow \eta $%
, $\varsigma _{I}\rightarrow \alpha $ ($-\alpha $), and $\varsigma
_{II}\rightarrow -\alpha $ ($\alpha $), which are satisfied when $\cos
\Omega t_{R}\cos \lambda t_{R}=1$. This relation implies that
\begin{equation}
t_{R}=%
{\displaystyle{n\pi  \over \lambda }}%
=\frac{m\pi }{\Omega },  \label{Eq54}
\end{equation}%
where $n,m$ are integers, both being even or odd. In Fig. 4(b), the
intermediate coupling $\lambda /\omega _{10}=1$ follows from $n=(4/5)m$,
such that the joint-system state recurs whenever
\begin{equation}
\lambda t_{R}=8n\pi ,n=0,1,2,...,  \label{Eq55}
\end{equation}%
a result which explains the shift in the scaled time $\lambda t$ observed in
Fig. 4(b).

Concerning the state-swap probability $P_{S}(t)$, we obtain the expression
\begin{eqnarray}
\left\langle \psi _{12}(0)|_{1(2)\rightarrow 2(1)}\right. \left| \psi
_{12}(t)\right\rangle &=&{\cal N}_{\pm }^{2}\left( \left\langle \alpha
\right. \left| \zeta _{I}\right\rangle \left\langle \eta \right. |\varsigma
_{I}\rangle +\exp \left( 2iG_{2}\alpha \sin \left( \lambda t\right) \right)
\left\langle -\alpha \right. \left| \zeta _{II}\right\rangle \left\langle
\eta \right. |\varsigma _{II}\rangle \right.  \nonumber \\
&&\left. \pm \left\langle -\alpha \right. \left| \zeta _{I}\right\rangle
\left\langle \eta \right. |\varsigma _{I}\rangle \pm \exp \left(
2iG_{2}\alpha \sin \left( \lambda t\right) \right) \left\langle \alpha
\right. \left| \zeta _{II}\right\rangle \left\langle \eta \right. |\varsigma
_{II}\rangle \right) ,  \label{Eq56}
\end{eqnarray}%
which tends to unity when $\varsigma _{I},\varsigma _{II}\rightarrow \eta $,
$\zeta _{I}\rightarrow \alpha $ ($-\alpha $), and $\zeta _{II}\rightarrow
-\alpha $ ($\alpha $). These conditions are satisfied only when $\sin \Omega
t_{S}\sin \lambda t_{S}=-1${\LARGE \ }and{\LARGE \ }$G_{1}$,$G_{2}\approx 0$%
. For the special case of $F=0$, as in Fig. 4(b and c), the condition $G_{1}$%
,$G_{2}\approx 0$ is automatically satisfied (as can be deduced from Eqs. (%
\ref{Eq36a}) and (\ref{Eq36b})) and the modes will swap their states
whenever
\begin{equation}
t_{S}=\frac{\left( 2n-1\right) \pi }{2\lambda }=\frac{\left( 2m+1\right) \pi
}{2\Omega }{\rm {,}}  \label{Eq57}
\end{equation}%
with the additional condition that $\sin \left[ \left( n-1/2\right) \pi %
\right] =-\sin \left[ \left( m+1/2\right) \pi \right] $. Therefore, the%
{\large \ }maxima of the swap probability are eliminated, as observed in
Fig. 4(b) (for the intermediate coupling $\lambda /\omega _{10}=1$) due to
the fact that the relation $2n-1=4\left( 2m+1\right) /5$ (and so $\sin
\Omega t_{S}\sin \lambda t_{S}=-1$) cannot be satisfied for any pair ($n,m$%
). However, owing to the oscillations of $P_{S}(t)$ within the envelop
function defined by $\lambda $($=4\Omega /5$), the neighborhood of these
maxima still survives. Differently, in Fig. 4(c) (for the strong coupling
regime $\Omega /\lambda =1$), the relation $\sin \Omega t_{S}\sin \lambda
t_{S}=\sin ^{2}\lambda t_{S}\neq -1$ implies that the single maximum of the
swap probability is eliminated (since $\Omega /\lambda =1$, there are no
oscillations of $P_{S}(t)$ but the envelope function). A heuristic
explanation of the behaviors of Figs. 4(b and c) will be provided below. In
fact, as discussed below, while the superposition state ${\cal N}_{\pm
}\left( |\alpha \rangle \pm |-\alpha \rangle \right) _{1}$ swaps to
oscillator $2$, the coherent state $|\eta \rangle _{2}$ does not swap to
oscillator $1$, even though both return to their respective systems.
Evidently, the phase mismatching between the coupling parameter $\lambda $
and the field-shifted frequencies $\omega _{\ell }=\Omega $ (when $F=0$),
represented by the relation $\sin \Omega t_{S}\sin \lambda t_{S}=\sin
^{2}\lambda t_{S}\neq -1$, is the core of this result.

In Fig. 4(d), also setting $F=0$, we include dissipation, taking both
systems with the same damping rate $\gamma _{\ell }^{\pm }=\Gamma /2$, where
$\Gamma /\omega _{10}=2\times 10^{-3}$, and $\lambda /\omega _{10}=2\times
10^{-2}$ (as in the weak coupling regime) and observing, as expected, the
relaxation of both probabilities $P_{R}(t)$ (solid line) and $P_{S}(t)$
(dotted line). To compare the relaxation of probabilities $P_{R}(t)$ and $%
P_{S}(t)$ in both regimes, in Fig. 4(e) we assume the parameters $F=0$, $%
\gamma _{\ell }^{+}=2\gamma _{\ell }^{-}=\Gamma /2$ (assuming Markovian
white noise$)$, $\Gamma /\omega _{10}=2\times 10^{-3}$, and $\lambda /\omega
_{10}=2$ (as in the strong coupling regime). We observe, comparing Figs.
4(d) and 4(e), that in the strong coupling regime the fields recur more
frequently, within the relaxation time, than in the weak coupling regime.
Owing to the strong coupling, in Fig. 4(e) the swap dynamics is almost
completely lost, as in Fig. 4(c). In Fig. 4(e), a dashed-dotted line has
been drawn at unity in order to display the slow decay of $P_{R}(t)$.

Finally, in Fig. 4(f) we again disregard dissipation, but turn on the
driving field, taking $F/\omega _{20}=1$ (i.e., $\omega _{10}=2\omega _{20}$%
), and $\lambda /\omega _{10}=2\times 10^{-2}$ (as in the weak coupling
regime), and observe a reduction of the swap probability compared to the
recurrence probability. (Note that the driving field occurs in the von
Neumann term of Eq. (\ref{Eq18}) even though it does not influence the
Liouville operator.) This behavior can be explained by the result
\begin{equation}
G_{1}-G_{2}=\frac{F}{\Omega -\lambda }{\rm {,}}  \label{Eq58}
\end{equation}%
which becomes $1/2$\ for the parameters considered above, preventing the
swap probability from being unity, as required by the conditions $\sin
\Omega t_{S}\sin \lambda t_{S}=-1${\LARGE \ }and{\LARGE \ }$%
G_{1}-G_{2}\approx 0$. We have assumed in this figure the ratio $\omega
/\omega _{10}=10^{-2}$, instead of the resonance condition $\omega /\omega
_{10}=1/2$ adopted above, in order to make clear the amplification effects.
>From Eq. (\ref{Eq56}) it follows immediately that the probability $P_{S}(t)$
is reduced by the factor $\exp \left\{ -2\left[ F/(\Omega -\lambda )\right]
^{2}\right\} $, which explains the maximum value around $1/2$ for the
probability $P_{S}(t)$. It is worth noting that the expression (\ref{Eq58})
does not diverge for $F\neq 0$, since in the case where $\Omega =\lambda $
(and switching off the dissipation), Eq. (\ref{Eq34l}) becomes

\begin{equation}
\frac{\partial ^{2}\alpha _{1}}{\partial t^{2}}-2\Omega \frac{\partial
\alpha _{1}}{\partial t}-iF\lambda =0,  \label{Eq58l}
\end{equation}
whose solution differs from that leading to Eq. (\ref{Eq58}).

It can be argued that the state-swap and recurrence dynamics are
consequences of energy transfer between modes, which in the presence of
dissipation is severely reduced, as the two modes tend to reach energy
equilibrium with the reservoirs and the driving field. Actually, state-swap
and recurrence are properties of information transfer rather than energy
transfer between systems, even though these quantities are generally
correlated. However, it can be shown \cite{meu} that for the coupling
between the modes selected above, even if each mode is kept to constant
energy, their states can be swapped, showing the independence of the two
processes. In the presence of the relaxation process, the information
transfer between modes is reduced, becouse of the absorption of information
by the reservoirs. When the systems are strongly interacting, however, the
field states recur many times on the scale of $\lambda $ before the
relaxation takes place. While not preventing the modes from relaxing to
equilibrium with the reservoirs, the recurrence has striking consequences
for short-time-scale dynamics, such as the dynamics of decoherence discussed
below.

\section{coherence properties}

So far we have analyzed the dynamics of strongly-interacting modes $1$ and $%
2 $ to discuss the recurrence and state swap processes. Now we analyze, also
in the strong coupling regime,\ with the driving field switched off ($F=0$)
and the reservoirs at absolute zero ($T=0$), the decoherence dynamics of the
joint state described by the density operator $\rho _{12}$, and of the state
of system $1$ ($2$), described by $\rho _{1}=%
\mathop{\rm Tr}%
_{2}\rho _{12}$ ($\rho _{2}=%
\mathop{\rm Tr}%
_{1}\rho _{12}$). In this section we also consider the case of identical
dissipative systems, $\gamma _{\ell }^{\pm }=\gamma ^{\pm }$. The case of
different decay rates will be analyzed subsequently. We consider three
different initial joint states: first, the disentangled state, given by Eq. (%
\ref{Eq46}), and then the entanglements which are eigenstates associated
with the normal modes $\omega _{\ell }^{\pm }$, derived from Eq. (\ref{Eq40}%
): with $\beta _{I}^{1}=\beta _{II}^{2}=\alpha $\ and $\beta _{I}^{2}=\beta
_{II}^{1}=-\alpha $ we obtain the eigenstate
\begin{equation}
\left| \varphi _{12}^{-}\right\rangle ={\cal N}_{\pm }\left( \left| \alpha
,-\alpha \right\rangle _{12}\pm \left| -\alpha ,\alpha \right\rangle
_{12}\right) {,}  \label{Eq59}
\end{equation}%
associated with the normal mode $\omega _{\ell }^{-}$, and with $\beta
_{I}^{1}=\beta _{I}^{2}=\alpha $\ and $\beta _{II}^{2}=\beta
_{II}^{1}=-\alpha $, we obtain
\begin{equation}
\left| \varphi _{12}^{+}\right\rangle ={\cal N}_{\pm }\left( \left| \alpha
,\alpha \right\rangle _{12}\pm \left| -\alpha ,-\alpha \right\rangle
_{12}\right) {,}  \label{Eq60}
\end{equation}%
which is the eigenstate associated with $\omega _{\ell }^{+}$.

\subsection{Decoherence time of the state{\it \ }$|\protect\psi _{12}\rangle
={}{\cal N}_{\pm }\left( |\protect\alpha \rangle \pm |-\protect\alpha %
\rangle \right) _{1}\otimes |\protect\eta \rangle _{2}$}

The coherence of the joint state (\ref{Eq46}) is given, essentially, by the
term
\begin{equation}
\exp \left\{ -2|\alpha |^{2}\left[ 1-\left( \left| W^{-}\right| ^{2}+\left|
Z^{-}\right| ^{2}\right) \exp \left( -2{\rm Re}\left\{ \Lambda \right\}
t\right) \right] \right\} ,  \label{Eq61}
\end{equation}%
coming from the off-diagonal coefficients of the density operator in Eq. (%
\ref{Eq47}). In the strong coupling regime, the exponential decay in Eq. (%
\ref{Eq61}), computed from Eqs. (\ref{Eq35a}), (\ref{Eq35b}), reduces to the
form
\begin{equation}
\exp \left[ -|\alpha |^{2}\left( 2-%
\mathop{\rm e}%
\nolimits^{-\left( \gamma _{1}^{+}+\gamma _{2}^{+}\right) t}-%
\mathop{\rm e}%
\nolimits^{-\left( \gamma _{1}^{-}+\gamma _{2}^{-}\right) t}\right) \right] {%
.}  \label{Eq62}
\end{equation}%
Assuming identical dissipative systems, $\gamma _{\ell }^{\pm }=\gamma ^{\pm
}$, Eq. (\ref{Eq62}) simplifies to
\begin{equation}
\exp \left[ -|\alpha |^{2}\left( 2-%
\mathop{\rm e}%
\nolimits^{-2\gamma ^{+}t}-%
\mathop{\rm e}%
\nolimits^{-2\gamma ^{-}t}\right) \right] {,}  \label{Eq63}
\end{equation}%
resulting in a decoherence time for the joint system given by
\begin{equation}
\tau _{D}=\left[ 2|\alpha |^{2}\left( \gamma ^{+}+\gamma ^{-}\right) \right]
^{-1}{\rm {.}}  \label{Eq63l}
\end{equation}%
This decoherence time has to be compared with that for an isolated mode in
the superposition state ${\cal N}_{\pm }\left( |\alpha \rangle \pm |-\alpha
\rangle \right) $ (also obtained from the weak coupling regime: $\gamma
^{\pm }=\Gamma /2$), following from the exponential decay
\begin{equation}
\exp \left[ -2|\alpha |^{2}\left( 1-%
\mathop{\rm e}%
\nolimits^{-\Gamma t}\right) \right]  \label{Eq64}
\end{equation}%
and given by the well-known expression $\tau _{D}\approx \tau _{R}/2|\alpha
|^{2}\equiv {\cal T}_{D}$, where $\tau _{R}=\Gamma ^{-1}$ stands for the
relaxation time of the system. Analyzing the decoherence process in the
light of the spectral densities considered in Section III, we observe that
for Markovian white noise ($M$), where $\gamma ^{-}=\gamma ^{+}/2=\Gamma /4$%
, the decoherence time in Eq. (\ref{Eq63l}) rises to
\begin{equation}
\tau _{D}^{M}\approx 4{\cal T}_{D}/3.  \label{EqTD1}
\end{equation}%
For the Lorentzian spectral density ($L$), where $\gamma ^{\pm }=\varepsilon
^{\pm }\Gamma /2\ll \Gamma $, we obtain the result
\begin{equation}
\tau _{D}^{L}\approx 2{\cal T}_{D}/(\varepsilon ^{+}+\varepsilon ^{-}),
\label{EqTD2}
\end{equation}%
which becomes large as the reservoir spectral density decreases around the
effective frequency $\omega _{\ell }^{\pm }=\omega ^{\pm }$. We stress that
for the case where $\varepsilon ^{+}=\varepsilon ^{-}$\ (i.e., $\gamma
^{+}=\gamma ^{-}$) the cross-decay channel is null.

Finally, for a wide Lorentzian spectral density ($WL$), where $\gamma
^{-}=\varepsilon ^{-}\Gamma /2\ll \Gamma $ and $\gamma ^{+}=\Gamma /2$, we
obtain from Eq. (\ref{Eq63l}), the value
\begin{equation}
\tau _{D}^{WL}\approx 2{\cal T}_{D}.  \label{EqTD3}
\end{equation}%
Therefore, for a wide Lorentzian spectral density we obtain a decoherence
time for strongly coupled systems which is twice as long as in the
weak-coupling regime. The mechanism behind these improved decoherence times
is that in the strong coupling regime (where the\ natural frequency $\omega
_{\ell 0}$ of the systems splits into two effective frequencies $\omega
_{\ell }^{\pm }=\omega _{\ell }\pm \lambda $) the spectral density of the
reservoir plays a decisive role in the damping rate, which also splits into
two Lorentzian functions. When the oscillator effective frequency $\omega
_{\ell }^{\pm }$ shifts to regions where the spectral density of the
reservoir is significantly smaller than that around $\omega _{\ell 0}$, the
damping rate becomes smaller than its value in the weak coupling regime.
Despite the spectral densities, the competition between the cross-decay and
the usual channels is the reason for the computed delay of the decoherence
process.

Still regarding state (\ref{Eq46}) and the strong coupling regime, focusing
on the reduced systems, we obtain from Eqs. (\ref{Eq45a}) and (\ref{Eq45b})
the density operators

\begin{eqnarray}
\rho _{1}(t) &=&{\cal N}_{\pm }^{2}\sum_{m,n=I,II}\left( \pm 1\right)
^{1-\delta _{mn}}\left\langle \beta _{n}^{1}\right. \left| \beta
_{m}^{1}\right\rangle ^{1-{\cal F}(\lambda t)/4}%
\mathop{\rm e}%
{}^{i\theta _{mn}^{(3)}+\theta _{mn}^{(4)}}\left| \varsigma
_{m}\right\rangle \left\langle \varsigma _{n}\right| ,  \label{Eq65a} \\
\rho _{2}(t) &=&{\cal N}_{\pm }^{2}\sum_{m,n=I,II}\left( \pm 1\right)
^{1-\delta _{mn}}\left\langle \beta _{n}^{1}\right. \left| \beta
_{m}^{1}\right\rangle ^{1-{\cal F}(\lambda t+\pi /2)/4}%
\mathop{\rm e}%
{}^{i\theta _{mn}^{(1)}+\theta _{mn}^{(2)}}\left| \zeta _{m}\right\rangle
\left\langle \zeta _{n}\right| ,  \label{Eq65b}
\end{eqnarray}%
where
\begin{equation}
{\cal F}(\lambda t)=\exp \left[ -(\gamma _{1}^{+}+\gamma _{2}^{+})t\right]
+\exp \left[ -(\gamma _{1}^{-}+\gamma _{2}^{-})t\right] +2\cos (2\lambda
t)\exp \left[ -(\gamma _{1}^{+}+\gamma _{2}^{+}+\gamma _{1}^{-}+\gamma
_{2}^{-})t/2\right] .  \label{Eq66}
\end{equation}%
Observe that for identical systems, $\gamma _{\ell }^{\pm }=\gamma ^{\pm }$,
the decoherence of the reduced state in oscillator$1$ is given by
\begin{equation}
\exp \left\{ -2|\alpha |^{2}\left[ 1-\frac{1}{4}\left( {\rm {%
\mathop{\rm e}%
}}^{-2\gamma ^{+}t}+{\rm {%
\mathop{\rm e}%
}}^{-2\gamma ^{-}t}+2\cos (2\lambda t){\rm {%
\mathop{\rm e}%
}}^{-(\gamma ^{+}+\gamma ^{-})t}\right) \right] \right\}  \label{Eq67a}
\end{equation}%
while for the state in oscillator $2$ it is
\begin{equation}
\exp \left\{ -2|\alpha |^{2}\left[ 1-\frac{1}{4}\left( {\rm {%
\mathop{\rm e}%
}}^{-2\gamma ^{+}t}+{\rm {%
\mathop{\rm e}%
}}^{-2\gamma ^{-}t}+2\cos (2\lambda t+\pi ){\rm {%
\mathop{\rm e}%
}}^{-(\gamma ^{+}+\gamma ^{-})t}\right) \right] \right\} .  \label{Eq67b}
\end{equation}%
It is easily shown that the expressions in Eqs. (\ref{Eq67a}) and (\ref%
{Eq67b}), associated with the decoherence times of the states of systems $1$
and $2$, respectively, oscillate below the curve for the coherence decay of
the joint system given by expression (\ref{Eq63}). In Fig. 5, assuming
Markovian white noise and setting the fictitious ratio $\lambda /\Gamma =5$
to make the oscillations clear, the dashed and dotted lines refer to the
decoherence dynamics of systems $1$ and $2$, dictated by Eqs. (\ref{Eq67a})
and (\ref{Eq67b}), respectively. The solid line represents the coherence
decay derived from Eq. (\ref{Eq63}) and the dashed-dotted line indicates the
coherence decay for an isolated mode, computed from Eq. (\ref{Eq64}).
Therefore, for the special case of the initially disentangled state given by
Eq. (\ref{Eq46}), the decoherence times of systems $1$ and $2$ coincide with
that of the joint system.

\subsection{Decoherence time of the state{\it \ }$\left| \protect\varphi %
_{12}^{-}\right\rangle ={\cal N}_{\pm }\left( \left| \protect\alpha ,-%
\protect\alpha \right\rangle _{12}\pm \left| -\protect\alpha ,\protect\alpha %
\right\rangle _{12}\right) $}

Next, we analyze the decoherence dynamics of the entangled state (\ref{Eq59}%
) for the joint system. The decoherence process of this joint state, in the
strong coupling regime, is given by the exponential decay
\begin{equation}
\exp \left[ -4\left| \alpha \right| ^{2}\left( 1-%
\mathop{\rm e}%
\nolimits^{-\left( \gamma _{1}^{-}+\gamma _{2}^{-}\right) t}\right) \right] ,
\label{Eq69}
\end{equation}%
which takes into account only the decay rate of the system-reservoir
coupling around the effective frequency $\omega _{\ell }^{-}$. At this point
it is interesting to identify the contribution of the cross-decay channel to
the decoherence process, rewriting Eq. (\ref{Eq69}) as
\begin{equation}
\exp \left\{ -4\left| \alpha \right| ^{2}\left( 1-\exp \left[ -%
\mathop{\textstyle\sum}%
\nolimits_{\ell }\left( \gamma _{\ell }^{+}+\gamma _{\ell }^{-}\right) t/2%
\right] \exp \left[
\mathop{\textstyle\sum}%
\nolimits_{\ell }\left( \gamma _{\ell }^{+}-\gamma _{\ell }^{-}\right) t/2%
\right] \right) \right\} .  \label{Eq(69l)}
\end{equation}%
In this expression, the term $\exp \left[
\mathop{\textstyle\sum}%
\nolimits_{\ell }\left( \gamma _{\ell }^{+}-\gamma _{\ell }^{-}\right) t/2%
\right] $ comes from the cross-decay channel and it is evident that for $%
\mathop{\textstyle\sum}%
\nolimits_{\ell }\left( \gamma _{\ell }^{+}-\gamma _{\ell }^{-}\right) >0$,
its contribution makes the exponential decay slower. Assuming identical
dissipative systems, $\gamma _{\ell }^{-}={\gamma }^{-}$, the expression (%
\ref{Eq69}) simplifies to
\begin{equation}
\exp \left[ -4\left| \alpha \right| ^{2}\left( 1-%
\mathop{\rm e}%
\nolimits^{-2\gamma ^{-}t}\right) \right] ,  \label{Eq70}
\end{equation}%
leading to the decoherence time $\left( 8\left| \alpha \right| ^{2}{\gamma }%
^{-}\right) ^{-1}$, which is ${\cal T}_{D}/2$ in the weak coupling regime,
as expected. However, in the strong coupling regime, this decoherence time
for an entangled state is equal to ${\cal T}_{D}$ for Markovian white noise
and is significantly improved for both Lorentzian spectral densities, where
we obtain ${\cal T}_{D}/2\varepsilon ^{-}$.

\subsection{Decoherence time of the state{\it \ }$\left| \protect\varphi %
_{12}^{+}\right\rangle ={\cal N}_{\pm }\left( \left| \protect\alpha ,\protect%
\alpha \right\rangle _{12}\pm \left| -\protect\alpha ,-\protect\alpha %
\right\rangle _{12}\right) $}

The decoherence process of the joint state $\left| \varphi
_{12}^{+}\right\rangle $, in the strong coupling regime, is given by the
exponential decay
\begin{equation}
\exp \left[ -4\left| \alpha \right| ^{2}\left( 1-%
\mathop{\rm e}%
\nolimits^{-\left( \gamma _{1}^{+}+\gamma _{2}^{+}\right) t}\right) \right] ,
\label{Eq71}
\end{equation}%
depending only on the decay rate of the system-reservoir coupling around the
effective frequency $\omega _{\ell }^{+}$. Rewriting Eq. (\ref{Eq71}) as
\begin{equation}
\exp \left\{ -4\left| \alpha \right| ^{2}\left( 1-\exp \left[ -%
\mathop{\textstyle\sum}%
\nolimits_{\ell }\left( \gamma _{\ell }^{+}+\gamma _{\ell }^{-}\right) t/2%
\right] \exp \left[ -%
\mathop{\textstyle\sum}%
\nolimits_{\ell }\left( \gamma _{\ell }^{+}-\gamma _{\ell }^{-}\right) t/2%
\right] \right) \right\} ,  \label{Eq71l}
\end{equation}%
we identify the term $\exp \left[ -%
\mathop{\textstyle\sum}%
\nolimits_{\ell }\left( \gamma _{\ell }^{+}-\gamma _{\ell }^{-}\right) t/2%
\right] $ as coming from the cross-decay channel. It is evident that \ for $%
\mathop{\textstyle\sum}%
\nolimits_{\ell }\left( \gamma _{\ell }^{+}-\gamma _{\ell }^{-}\right) >0$,
its contribution speeds up the exponential decay, in contrast to the above
situation where the eigenstate of the normal mode $\omega _{\ell }^{-}$ is
considered. For $\gamma _{\ell }^{-}={\gamma }^{-}$, Eq. (\ref{Eq71})
simplifies to
\begin{equation}
\exp \left[ -4\left| \alpha \right| ^{2}\left( 1-%
\mathop{\rm e}%
\nolimits^{-2\gamma ^{+}t}\right) \right] ,  \label{Eq72}
\end{equation}%
and the decoherence time $\left( 8\left| \alpha \right| ^{2}{\gamma }%
^{+}\right) ^{-1}$ becomes ${\cal T}_{D}/2$ for the weak coupling regime and
also for both spectral densities, Markovian white noise and the wide
Lorentzian spectral noise, in the strong coupling regime. For the Lorentzian
spectral density, we obtain ${\cal T}_{D}/2\varepsilon ^{+}$. In the light
of the above results we conclude, apart from the dependence of the
decoherence time upon the reservoir spectral densities, that the eigenstate
associated with the normal mode $\omega _{\ell }^{-}$ is less susceptible to
decoherence than that associated with $\omega _{\ell }^{+}$.

\section{Improving the quality factor of a dissipative system}

Let us now turn to a central result, which is the effect of the coupling $%
\lambda $ on the decoherence process for states of oscillators with
different damping rates $\Gamma _{1}$ and $\Gamma _{2}$. We assume that the
state of the whole system is given by Eq. (\ref{Eq46}) and that oscillator $%
2 $ has a better quality factor than oscillator $1$ (i.e., $\Gamma
_{1}=\Gamma \gg \Gamma _{2}$), where we have the superposition ${\cal N}%
_{\pm }\left( |\alpha \rangle \pm |-\alpha \rangle \right) _{1}$ which we
want to protect. The decoherence of state $1$, following from Eq. (\ref%
{Eq65a}), is then governed by the exponential decay
\begin{equation}
\exp \left\{ -2|\alpha |^{2}\left[ 1-\frac{1}{4}\left(
\mathop{\rm e}%
\nolimits^{-\gamma ^{-}t}+%
\mathop{\rm e}%
\nolimits^{-\gamma ^{+}t}+2\cos (2\lambda t)%
\mathop{\rm e}%
\nolimits^{-\left( \gamma ^{+}+\gamma ^{-}\right) t/2}\right) \right]
\right\}  \label{Eq74a}
\end{equation}%
while for the field state $2$, following from (\ref{Eq65b}), it is governed
by
\begin{equation}
\exp \left\{ -2|\alpha |^{2}\left[ 1-\frac{1}{4}\left(
\mathop{\rm e}%
\nolimits^{-\gamma ^{-}t}+%
\mathop{\rm e}%
\nolimits^{-\gamma ^{+}t}+2\cos (2\lambda t+\pi )%
\mathop{\rm e}%
\nolimits^{-\left( \gamma ^{+}+\gamma ^{-}\right) t/2}\right) \right]
\right\} .  \label{Eq74b}
\end{equation}%
These expressions are to be compared with Eqs. (\ref{Eq67a}) and (\ref{Eq67b}%
), respectively, where we have assumed the same quality factors for both
oscillators. The decoherence times obtained from Eqs. (\ref{Eq74a}) and (\ref%
{Eq74b}) have an upper limit given by the exponential decay for the joint
system
\begin{equation}
\exp \left[ -|\alpha |^{2}\left( 2-%
\mathop{\rm e}%
\nolimits^{-\gamma ^{+}t}-%
\mathop{\rm e}%
\nolimits^{-\gamma ^{-}t}\right) \right] ,  \label{Eq75}
\end{equation}%
which is derived from Eq. (\ref{Eq62}). This exponential decay leads to a
decoherence time given by
\begin{equation}
\tau _{D}\approx \left[ |\alpha |^{2}\left( \gamma ^{+}+\gamma ^{-}\right) %
\right] ^{-1},  \label{Eq76}
\end{equation}%
which is twice the result found in Eq. (\ref{Eq63l}), for the case of
identical dissipative systems ($\gamma _{\ell }^{\pm }=\gamma ^{\pm }$),
whatever the spectral density of the reservoir. Therefore, even in the weak
coupling regime, the decoherence time of a system is improved when it is
coupled to another system with a better quality factor. We stress that this
conclusion holds only when $\lambda \gg \gamma _{\ell }^{\pm }$, a situation
we have assumed even for the weak coupling regime. Evidently, when $\lambda
\lessapprox \gamma _{\ell }^{\pm }$, so that the recurrence-swap dynamics
does not take place effectively before the relaxation time, the quality
factor of a system cannot be improved by coupling it to another system of
better quality.

In Fig. 6, assuming the weak coupling regime, $\gamma _{\ell }^{\pm }=\Gamma
_{\ell }/2$, with $\Gamma _{1}/\Gamma _{2}=10^{2}$ and $\lambda /\Gamma
_{1}=5$ (a ratio chosen to show the dissipative dynamics clearly), the
dashed and dotted lines refer to the decoherence dynamics of systems $1$ and
$2$, derived from Eqs. (\ref{Eq74a}) and (\ref{Eq74b}), respectively. The
solid line represents the coherence decay for the joint system derived from
Eq. (\ref{Eq75}) and the dashed-dotted line indicates the coherence decay
for the isolated mode $1$, where the superpositon state ${\cal N}_{\pm
}\left( |\alpha \rangle \pm |-\alpha \rangle \right) _{1}$ is prepared,
computed from Eq. (\ref{Eq64}). However, while the bad-quality system $1$
gets better when coupled to a good-quality system $2$, the latter, in turn,
gets worse. In fact, the thick solid line in Fig. 6, representing the
decoherence process of the superposition ${\cal N}_{\pm }\left( |\alpha
\rangle \pm |-\alpha \rangle \right) _{2}$ prepared in an isolated
good-quality system $2$, displays a slower decay rate than the solid line
when system $2$ is coupled to a bad-quality system $1$. In conclusion, in a
network made of two oscillators with different quality factors, the
good-quality oscillator gets worse, while the bad-quality oscillator gets
better. We hope that this process may be even improved by extending the
coupling of a bad-quality oscillator to a higher number of good-quality
oscillators.

Finally, we note that the above result can be theoretically employed to
protect quantum superposition states generated, via atom-field interaction,
in open cavities. By coupling an open bad-quality cavity, where we have
prepared a quantum superposition, to a closed good-quality cavity, we can
protect the superposition through the coupling between the cavities and,
further, in an appropriate recurrence time, we can recover that
superposition state in the open cavity. The process works as if the original
superposition coherence is protected from the dissipative mechanism by the
recurrence-swap dynamics, much like a system which is put in contact with a
high-temperature reservoir, but is intermittently brought into contact with
a low-temperature reservoir. If the interval between contacts is fast
enough, the system will take longer to relax to the hot reservoir
temperature than it would in the absence of the cold reservoir. Of course,
this analogy has to be considered cautiously, since temperature and
coherence are very different features.

In the next section, we discuss how the original superposition state can be
recovered in the bad-quality system after its coupling with the good-quality
system.

\section{Entropy Excess}

To clarify the role the state-swap processes play in the coherence dynamics,
we consider the superposition state in Eq. (\ref{Eq46}), $|\psi _{12}\rangle
={}{\cal N}_{\pm }\left( |\alpha \rangle \pm |-\alpha \rangle \right)
_{1}\otimes |\eta \rangle _{2}$, and plot, in Fig. 7(a and b), the
probabilities
\begin{equation}
{\cal P}_{\ell }(t)=\left\langle \psi _{m}\left| \rho _{\ell }(t)\right|
\psi _{m}\right\rangle ,  \label{Eq77}
\end{equation}%
of finding the superposition state $|\psi _{1}\rangle ={}{\cal N}_{\pm
}\left( |\alpha \rangle \pm |-\alpha \rangle \right) _{1}$ and the coherent
state $|\psi _{2}\rangle =|\eta \rangle _{2}$ in oscillator $\ell $. In
these figures we assume the strong coupling regime ($\lambda /\omega _{10}=2$%
), identical dissipative systems ($\gamma _{\ell }^{\pm }=\gamma ^{\pm }$),
and Markovian white noise ($\gamma ^{+}=2\gamma ^{-}=\Gamma /2$). The ratio $%
\lambda /\Gamma =20$ is set to a fictitious scale to make clear the
state-swap and state-recurrence dynamics, as the strong oscillations
obtained with a realistic $\lambda $ would be blurred by the slow
(dissipative) dynamics.

>From the solid line in Fig. 7(a), we observe that the superposition $|\psi
_{1}\rangle $ recurs to oscillator $1$, despite the dissipative process.
However, the coherent state $|\eta \rangle _{2}$ does not swap to oscillator
$1,$ as indicated by the dotted line. In fact, the swapping of \ the
coherent state $|\eta \rangle _{2}$ to oscillator $1$ would be indicated by
the occurrence of maxima of the dotted line between those of the solid line.
In Fig. (b) we observe that the coherent state $|\eta \rangle _{2}$ recurs
to oscillator $2$, as indicated by the solid line. Moreover, the state $%
|\psi _{1}\rangle $ also swaps to oscillator $2$ (dotted line), indicating
that the superposition ${\cal N}_{\pm }\left( |\alpha \rangle \pm |-\alpha
\rangle \right) $ is completely interchanged between the systems,
differently from the coherent state which does not swap to oscillator $1$.

Therefore, from Fig. 7(a) we conclude that the superposition state ${\cal N}%
_{\pm }\left( |\alpha \rangle \pm |-\alpha \rangle \right) $ prepared in a
bad-quality system and protected from decoherence by coupling this system to
a good-quality one, can easily be recovered in system $1$ by switching off
the coupling at the recurrence time $t_{R}=n\pi /\lambda ,{\rm {\ }}%
n=0,1,2,....$ After this time, the superposition state will be in oscillator
$1$ with a fidelity less than unity due to the dissipative process.

In Figs. 8(a and b) we plot the linear entropy for the joint state in Eq. (%
\ref{Eq46}) (${\cal S}_{12}=1-$ {\rm Tr}$\rho _{12}^{2}$) and the reduced
states of oscillators $1$ and $2$ (${\cal S}_{\ell }=1-$ {\rm Tr}$\rho
_{\ell }^{2}$). In these figures we employ the same parameters considered in
Fig. 7, except that in Fig. 8(b) we use the ratio $\lambda /\Gamma =2$,
instead of $\lambda /\Gamma =20$, to show clearly the dissipative dynamics.
In Fig. 8(a) we analyze the recurrence-swap dynamics until around the
correlation time $\tau _{C}$, defined as the time when the entropy ${\cal S}%
_{1}$(${\cal S}_{2}$) goes to about $0.1$ at the swap (recurrence) time. In
fact, as shown in Fig. 8(a), the minima of the entropy ${\cal S}_{\ell }$
move away from zero due to the development of an inevitable correlation
between the oscillators (due to the cross-decay channel) which thus become
permanently entangled. This correlation time, estimated as the time when the
minima of ${\cal S}_{\ell }$ approach $0.1$, is given by
\begin{equation}
\tau _{C}\approx \frac{1}{5\left| \alpha \right| \sum_{\ell }\left( \gamma
_{\ell }^{+}-\gamma _{\ell }^{-}\right) }.  \label{Eq78}
\end{equation}%
>From Eq. (\ref{Eq78}) we conclude that for the weak coupling regime, where $%
\gamma _{\ell }^{+}=\gamma _{\ell }^{-}$, the correlation time goes to
infinity, i.e., the entropy ${\cal S}_{1}$(${\cal S}_{2}$) always returns to
zero in the swap (recurrence) time. Therefore, in the weak coupling regime
the oscillators do not get permanently entangled (due to the absence of the
cross-decay channel) and, since ${\cal S}_{1}$(${\cal S}_{2}$) always
returns to zero in the swap (recurrence) time, one is always able to\
recover a superposition state of a bad-quality oscillator coupled to a
good-quality one. However, even in the strong coupling regime the
correlation developed between the oscillators could not affect the process
of recovering a superposition state of a bad-quality oscillator coupled to a
good-quality one. Since the ratio $\tau _{C}/\tau _{D}$ obeys
\begin{equation}
\frac{\tau _{C}}{\tau _{D}}\approx \frac{\left| \alpha \right| }{5}\frac{%
\sum_{\ell }\left( \gamma _{\ell }^{+}+\gamma _{\ell }^{-}\right) }{%
\sum_{\ell }\left( \gamma _{\ell }^{+}-\gamma _{\ell }^{-}\right) },
\label{Eq79}
\end{equation}%
for $\tau _{C}/\tau _{D}\gtrsim 1$ one can always recover the superposition
state ${\cal N}_{\pm }\left( |\alpha \rangle \pm |-\alpha \rangle \right) $,
with a considerable fidelity, in spite of the process of entanglement
between the oscillators.

The overall picture coming from Fig. 8(a) is that the recurrence dynamics
due to the strong coupling between the oscillators tends to restore the
coherence of the initial states while the dissipative dynamics promotes the
decoherence process. In Fig. 8(b) the linear entropy of the joint state $%
{\cal S}_{12}$, represented by the thick solid line, starts from zero, goes
to a maximum due to the decoherence process and then returns to zero, since
in the asymptotic limit both oscillators reach a pure state: the vacuum or
some coherent state whose excitation depends on the amplification parameter $%
F$. Meanwhile, as shown in Fig. 8(a), the linear entropy ${\cal S}_{\ell }$
of the reduced state of oscillator $\ell $, oscillates between zero and $0.5$%
. The linear entropy ${\cal S}_{1}$(${\cal S}_{2}$), solid (dashed) line,
becomes zero when oscillator $1$($2$)\ assumes the state $|-\eta \rangle
_{1} $($|\eta \rangle _{2}$), as can be computed from Eqs. (\ref{Eq43a}) and
(\ref{Eq43b}).{\LARGE \ }At the same time, ${\cal S}_{2}$(${\cal S}_{1}$)
bump into the thick solid line representing the linear entropy for the joint
state of the system ${\cal S}_{12}$, from above, indicating that the
superposition $|\psi _{1}\rangle $ has swapped (recurred) to oscillator $2$($%
1$) on its way to decoherence. The maximal correlation between fields occurs
at the points where the recurrence and swap curves cross, as illustrated by
the dotted line representing the excess entropy, defined as
\begin{equation}
{\cal I}\equiv {\cal S}_{1}+{\cal S}_{2}-{\cal S}_{12}.  \label{Eq80}
\end{equation}%
We also observe from the dotted line that the minima of the excess entropy $%
{\cal I}$ move away from zero, due to the development of an inevitable
correlation between the oscillators. Returning to Fig. 8(b), after reaching
its maximum the correlation ${\cal S}_{12}$ decays exponentially as a result
of the dissipation and the driving field, attaining zero in the asymptotic
limit when there is no correlation between the fields described by
stationary factorized coherent states. During the decay of correlation $%
{\cal S}_{12}$, the linear entropy ${\cal S}_{\ell }$ of the reduced state
of oscillator $\ell $ does not attain the value ${\cal S}_{12}$, due to the
correlation described by the excess entropy ${\cal I}$.

\section{Conclusion}

We have presented a comprehensive treatment of the coherence dynamics in a
network composed of two coupled dissipative oscillators. First, we have
derived a master equation for both regimes of weak and strong coupling
between the oscillators. In the weak coupling regime the dissipative
mechanism of the individual oscillators is not significantly affected by
their interaction, which appears only in the von Neumann term of the master
equation. However, in the strong coupling regime the time evolution of the
density operator of the joint system is modified by a cross-decay channel
represented by a Liouville operator ${\cal L}_{12}\rho _{12}$ accounting for
the coupling between the oscillators. The appearance of this cross-decay
channel leads to interesting properties of the coherence dynamics of
strongly coupled oscillators.

After the mathematical development, we first analyzed the state-swap and the
state-recurrence dynamics, i.e., the probability that each oscillator
returns to its initial state and the probability of state-swapping between
the oscillators, respectively. In particular, we have analyzed these
processes for the case where both oscillators, prepared in the joint state $%
{\cal N}_{\pm }\left( |\alpha \rangle \pm |-\alpha \rangle \right)
_{1}\otimes |\eta \rangle _{2}$, present the same quality factor $\Gamma $.
In the weak coupling regime the recurrence and swap processes are fully
accomplished apart from the relaxation of the field states due to
dissipation. On the way to the strong coupling regime the recurrence process
remains unchanged while the swap dynamics is gradually lost as the coupling
gets stronger. In the strong coupling regime the swap dynamics is lost due
to phase mismatching between the coupling parameter $\lambda $ and the
field-shifted frequencies $\omega _{\ell }$, as explained in Sec. V.
Evidently, in the strong coupling regime the field states recur much more
often than in the weak coupling regime before the relaxation takes place.

Next, aware of the decoherence dynamics in the weak coupling regime,
governed by the usual master equation (\ref{Eq19}) where a Liouville
operator ${\cal L}_{\ell }\rho _{12}$ accounts for the effect of the
reservoir on oscillator $\ell $, we turn to the decoherence process in the
strong coupling regime. In this regime, the normal-mode frequencies ($\omega
_{\ell }^{\pm }=\omega _{\ell }\pm \lambda $) of the coupled systems are
strongly shifted, away from the oscillator frequencies $\omega _{\ell 0}$,
to regions of the reservoir frequency spaces where the spectral densities
may be significantly different from that around $\omega _{\ell 0}$. As the
spectral densities of the reservoirs play an important role in this regime,
three different spectral functions were considered for our analysis of the
decoherence process. When the normal-mode frequency $\omega _{\ell }^{-}$ is
shifted to regions around the origin of the frequency space, even for
Markovian white noise the coupling parameter of oscillator $\ell $ with its
reservoir, around $\omega _{\ell }^{-}$, becomes half its value around $%
\omega _{\ell }^{+}$ (i.e., $\gamma _{\ell }^{-}=\Gamma /4$). When
considering, instead, a Lorentzian spectral function, which approaches zero
around the origin of the reservoir frequency space, the coupling constant $%
\gamma _{\ell }^{-}$ becomes even smaller, resulting in decoherence times
for the joint and the reduced system states significantly longer than the
result computed for a single system plus reservoir. In Sec. VI the
decoherence times for three different states of the composed system are
computed, considering the three distinct reservoir spectral functions. Apart
from the dependence of the decoherence time upon the reservoir spectral
density, we stress that the coherence decay of the eigenstate of the normal
mode $\omega _{\ell }^{-}$ becomes slower than that for the eigenstate of
the normal mode $\omega _{\ell }^{+}$.

As discussed in Sec. III, apart from the possibility of considering
particular physical systems with appropriate reservoir spectral densities,
it is possible that specific spectral functions could be achieved through
engineered reservoirs. The difficult task of engineering strong interactions
between the oscillators, together with the achievement of specific reservoir
spectral functions, are the most sensitive problems in the way of the
physical implementation of the network here proposed. However, this proposal
might provide a motivation for future theoretical and experimental
investigations. In particular, the proposal could be applied to test the
Markovian white noise approximation or to probe the reservoir spectral
functions.

When considering the two systems to have different damping rates $\Gamma
_{1} $ and $\Gamma _{2}$ (with $\lambda \gg \Gamma _{1}$,$\Gamma _{2}$) we
demonstrated that the coupling between the systems, independently of its
strength, makes the good-quality system worse and the bad-quality system
better. This result can be employed to improve the quality factor of a
cavity, and thus to protect quantum superposition states generated, via
atom-field interaction, in open bad-quality cavities coupled to closed
good-quality ones. Evidently, this results holds for the weak coupling
regime because we have assumed that the coupling strength $\lambda $ is
significantly larger than the system damping rates even for that regime. In
fact, when considering $\lambda \gg \Gamma _{\ell }$, the computed
improvement of the quality factor of a system follows from the
recurrence-swap dynamics which take place many times before the relaxation
time, protecting in a good-quality system a field state originally prepared
in a bad-quality one.

Finally, we have developed a careful analysis of the entropy excess in our
network. Supposing that the joint system is prepared in the factorized state
$|\psi _{12}\rangle ={}{\cal N}_{\pm }\left( |\alpha \rangle \pm |-\alpha
\rangle \right) _{1}\otimes |\eta \rangle _{2}${,} we observed that in the
weak-coupling regime the recurrence-swap dynamics take place uninterruptedly
until the relaxation process rules it out, i.e., the entropy excess always
returns to zero in the recurrence and swap times. Therefore, in the
weak-coupling regime both oscillators always get disentangled in the
recurrence and swap times. Differently, in the strong coupling regime a
correlation is developed between the states ${\cal N}_{\pm }\left( |\alpha
\rangle \pm |-\alpha \rangle \right) $ and $|\eta \rangle $, in a such way
that they get permanently entangled after a time interval we have called the
correlation time. In this way, one cannot recuperate - with a fidelity equal
to unity - a superposition state ${\cal N}_{\pm }\left( |\alpha \rangle \pm
|-\alpha \rangle \right) $ prepared in a bad-quality system and protected
from decoherence through a strong coupling of such system with a
good-quality one. However, we computed the amplitude $\left| \alpha \right| $
where the correlation time becomes longer than the decoherence time,
allowing the recuperation of the superposition state ${\cal N}_{\pm }\left(
|\alpha \rangle \pm |-\alpha \rangle \right) $ with a good fidelity.

One of the main restrictions argued against quantum computation is that even
if each individual logic unit were only slightly affected by the decoherence
process, the coupling of a large number of such logic cells, for an actual
implementation such as the factorization of large numbers \cite{nielsen,shor}%
, would decrease the decoherence time in such a way that the whole computing
process would be seriously compromised \cite{Haroche1}. Here we have shown,
at least for two strongly coupled sites, that the decoherence process is not
dictated simply by the excitation of the state involved in the logic
operation and the damping decay rate of the logic cells. In a strongly
interacting quantum network the decoherence process is delayed by the
cross-decay channel arising from the strong coupling between the logic
cells. In fact it may occur that for many coupled sites, as for the two
coupled cavities presented in this paper, the decoherence time increases.
Therefore, the discussion about strongly interacting oscillators we have
presented here is central to understand the decoherence process in a quantum
network; specifically, our conclusion is that the decoherence time depends
not only on the excitation of the entanglement involved in the logic
operation but also on the coupling strength between the logic units.

$^{1}$maponte@df.ufscar.br

$^{2}$marcos@df.ufscar.br

$^{3}$miled@df.ufscar.br

{\bf Acknowledgments}

We wish to express thanks for the support from FAPESP (under contracts
\#99/11859-3, \#00/15084-5, and \#02/02633-6) and CNPq (Intituto do Mil\^{e}%
nio de Informa\c{c}\~{a}o Qu\^{a}ntica), Brazilian agencies. We also thank
R. M. Serra, C. J. Villas-B\^{o}as, V. V. Dodonov, and A. F. R. de Toledo
Piza for helpful discussions.

\newpage {\bf figures caption}

\bigskip

Fig. 1 Sketch of the coupled dissipative oscillators, with oscillator $2$
submitted to a classical driving field.

Fig. 2 Damping function $\gamma _{\ell }\left( \chi \right) $ assuming a
Lorentzian coupling $V_{\ell }$ between oscillator $\ell $ and its
respective reservoir. In the weak coupling regime the function $\gamma
_{\ell }\left( \chi \right) $ is centered around $\omega _{\ell }$ (dotted
line). As $\lambda $ increases, the damping function splits into two
Lorentzian functions whose peak heights are half the original value $\Gamma
_{\ell }$ (solid line). On the way to the strong coupling regime the two
peaks can be clearly distinguished as shown by the dashed line.

Fig. 3 Spectral density of the reservoir $\sigma _{\ell }\left( \chi \right)
$ for (a) Markovian white noise, (b) a Lorentzian spectral density, and (c)
a wide Lorentzian spectral density. The system-reservoir couplings around $%
\omega _{\ell }^{\pm }$ are represent by the shaded regions.

Fig. 4{\bf \ }State-swap probability $P_{S}(t)$ (dotted line) and recurrence
probability $P_{R}(t)$ (solid line) as a function of the scaled time $%
\lambda t$, for the factorized state\ $|\psi _{12}\rangle ={}{\cal N}_{\pm
}\left( |\alpha \rangle \pm |-\alpha \rangle \right) _{1}\otimes |\eta
\rangle _{2}$, taking $\alpha =\eta =1$ as real parameters. In (a)-(c)
relaxation and driving field are disregarded and we assume (a) weak
coupling, (b) an intermediate coupling, and (c) strong coupling regimes. In
(d) and (e) we set $F=0$ but dissipation is included, considering (d) weak
coupling and (e) strong coupling regimes. Finally, in (f) dissipation is
disregarded and the driving field turned on.

Fig. 5 Decoherence dynamics for the factorized state $|\psi _{12}\rangle ={}%
{\cal N}_{\pm }\left( |\alpha \rangle \pm |-\alpha \rangle \right)
_{1}\otimes |\eta \rangle _{2}$, assuming both oscillators with the same
damping factor. Coherence decays of oscillator $1$ (dashed line), oscillator
$2$ (dotted line), joint system (solid line) and an isolated oscillator with
damping rate $\Gamma _{1}$ (dashed-dotted line) are indicated.

Fig. 6 Decoherence dynamics for the factorized state $|\psi _{12}\rangle ={}%
{\cal N}_{\pm }\left( |\alpha \rangle \pm |-\alpha \rangle \right)
_{1}\otimes |\eta \rangle _{2}$, assuming that the damping rate of
oscillator $1$ is larger than that of oscillator $2$. The dashed (dotted)
line refers to the decoherence dynamics of oscillator $1$($2$), \ while the
solid (dashed-dotted) line represents the coherence decay of the joint
system (an isolated system with the damping rate of oscillator $1$),
respectively. The thick solid line represents the coherence decay of a
superposition state ${\cal N}_{\pm }\left( |\alpha \rangle \pm |-\alpha
\rangle \right) _{2}$ prepared in an isolated oscillator $2$.

Fig. 7 Probability ${\cal P}_{\ell }(t)=\left\langle \psi _{m}\left| \rho
_{\ell }(t)\right| \psi _{m}\right\rangle $, of finding the superposition
state $|\psi _{1}\rangle ={}{\cal N}_{\pm }\left( |\alpha \rangle \pm
|-\alpha \rangle \right) _{1}$ and the coherent state $|\psi _{2}\rangle
=|\eta \rangle _{2}$ in oscillator $\ell $. We consider (a) $\ell =1$ and
(b) $\ell =2$. In these figures the strong coupling regime is assumed
together with identical dissipative systems and Markovian white noise.

Fig. 8 Linear entropy for the joint state (${\cal S}_{12}=1-$ {\rm Tr}$\rho
_{12}$), represented by the thick solid line, and the reduced states of
oscillators $1$ and $2$ (${\cal S}_{\ell }=1-$ {\rm Tr}$\rho _{\ell }$),
represented by solid and dashed lines, respectively. The dotted line
indicates the excess entropy. In Fig. 8(a) the recurrence-swap dynamics is
plotted until around the correlation time $\tau _{C}$ when a permanent
correlation is developed between the oscillators. In Fig. 8(b) we are
concerned with the relaxation of the network.

\end{document}